\documentclass{aa}

\usepackage{graphicx}
\usepackage{txfonts}
\usepackage{xcolor}
\usepackage{hyperref}
\usepackage{placeins}

\begin{document}

\title{Coevolution of halo and quasar properties in dense environments: CARLA J1017+6116 at z=2.8}

\author{Sofia G. Gallego
        \inst{1}
        \and
        Simona Mei \inst{1,2,3}
        \and
        Christopher Martin \inst{3}
        \and
        Donal O'Sullivan \inst{3}
        \and
        Emanuele Daddi \inst{4}
        \and
        Dominika Wylezalek \inst{5}
        \and
        Nicholas Seymour \inst{6}
    }

\institute{
    Universit\'e Paris Cit\'e, CNRS(/IN2P3), Astroparticule et Cosmologie, F-75013 Paris, France\\
    \email{sofiag.gallego@gmail.com}
    \and 
    CNRS-UCB International Research Laboratory, Centre Pierre Binétruy, IRL2007, CPB-IN2P3, Berkeley, USA
    \and
    Cahill Center for Astrophysics, California Institute of Technology, 1216 East California Boulevard, Mail code 278-17, Pasadena, CA 91125, USA
\and
    CEA, Irfu, DAp, AIM, Université Paris-Saclay, Université de Paris, CNRS, 91191 Gif-sur-Yvette, France
\and Zentrum f\"ur Astronomie der Universit\"at Heidelberg, Astronomisches Rechen-Institut, M\"onchhofstr 12-14, D-69120 Heidelberg, Germany
\and
International Center for Radio Astronomy Research, Curtin University, GPO Box U1987, 6102 Perth, Australia
}

\date{}

\abstract
{Radio-loud active galactic nuclei, in particular radio-loud quasars, are fueled by accretion onto supermassive black holes and are among the most energetic sources in the Universe.
While their impact on their surroundings — from the interstellar medium to the circumgalactic medium — is well recognized, the specific mechanisms remain uncertain. 
In this study we analyze deep Keck Cosmic Web Imager observations of the Ly$\alpha$ halo surrounding the radio-loud quasar at the center of the cluster CARLA J1017+6116 at $z = 2.8$.  As is known from previous observations, the cluster hosts a high fraction of early-type galaxies, and the star formation of its spectroscopically confirmed cluster members is typical  of or higher than that of galaxies on the main sequence.
We find that the Ly$\alpha$ halo extends at least 16\arcsec\ (128 physical kpc) to a level of surface brightness of $10^{-19}$ $\mathrm{erg\,s^{-1}\,cm^{-2}\,arcsec^{-2}}$, with a total observed Ly$\alpha$ luminosity of ${43.35\pm0.05}\,\rm{log_{10}\,L_\odot}$. 
The halo has distinct kinematic regions with asymmetries suggestive of complex interactions between the quasar and the intracluster medium, possibly driven by a combination of biconical feedback and episodic activity. 
Despite the quasar classification, our reanalysis of very long baseline interferometry data finds no evidence of extended jet structures; we instead find compact and variable radio emission that could indicate episodic jet activity or suppression by the dense interstellar medium. 
Combining these observations with imaging obtained with the \textit{Hubble} Space Telescope, we identified one Ly$\alpha$-emitting source within the quasar halo. 
While mechanical feedback from a jet
appears limited or episodic, radiative feedback likely plays a dominant role in shaping the extended Ly$\alpha$ halo, highlighting the complex interplay between quasar-driven processes and the surrounding dense environment.
}


\maketitle

\section{Introduction}

The formation and evolution of galaxies are heavily influenced by their environments and the various astrophysical processes occurring on both small and large scales. On smaller scales, processes such as star formation, supernova feedback, active galactic nucleus (AGN) feedback, and the dynamics of gas within individual galaxies are crucial \citep{1998ApJ...498..541K,2014ARA&A..52..415M, annurev:/content/journals/10.1146/annurev-astro-081811-125521, 2015MNRAS.446..521S, 2017FrASS...4...42M, 2021Univ....7..142E}. On larger scales, phenomena like galaxy mergers \citep{1993MNRAS.262..627L, 2011ApJ...730....4B, 2012A&A...539A..45L}, tidal interactions within groups and clusters \citep{1994ApJ...427..696M, 2003ApJ...582..141G, 2011ApJ...739L..33C}, and the accretion of matter along cosmic filaments play significant roles \citep{2003ASSL..281..185K,2005MNRAS.363....2K, 2012ARA&A..50..491P, 2015MNRAS.446..521S, 2024MNRAS.534.1682O}. While we have increasingly detailed knowledge of the physical processes driving galaxy formation and evolution at the galactic level \citep{2000MNRAS.319..168C, 2005Natur.435..629S, 2005MNRAS.362..799M, 2014ARA&A..52..415M, 2018MNRAS.473.4077P}, and we understand the formation of the cosmic web -- a vast network of interconnected filaments, voids, and nodes that make up the large-scale structure of the Universe \citep{Peebles1975, 1996Natur.380..603B, vandeWeygaert2008} -- the interplay between these large-scale structures and smaller-scale processes remains an area of active investigation \citep{1984Natur.311..517B, 2011piim.book.....D,2013MNRAS.435..999D,2015MNRAS.448.3665E,2019ComAC...6....2N,2022MNRAS.510..581K,2023ApJ...954...31C,2024ApJ...970..177H}.

Central to this discussion are extremely luminous AGNs powered by supermassive black holes, including quasars, radio-loud AGNs (RLAGNs), and blazars. RLAGNs, in particular, are often found in overdense environments at high redshifts, such as galaxy clusters and protoclusters \citep{Venemans2007,2013ApJ...769...79W,2014MNRAS.445..280H,2017ApJ...846L..31D}, where their feedback effects are amplified, potentially influencing the evolution of multiple neighboring galaxies. They profoundly impact their surroundings through powerful outflows and intense radiation, influencing the formation and evolution of galaxies over intergalactic scales \citep{2005ARA&A..43..769V, 2014ARA&A..52..589H}. These outflows, often biconical in shape \citep{2000ApJ...545...63E, 2023Natur.624...53G}, can extend well beyond the host galaxy, affecting the circumgalactic medium (CGM) and potentially the entire cluster and protocluster environment. Studying RLAGN outflows provides crucial insights into the mechanisms driving galaxy evolution and the interplay between supermassive black holes and their host galaxies.

The impact of RLAGN-driven outflows is particularly relevant at redshifts $2<z<3$, a crucial epoch in cosmic history often termed cosmic noon, when star formation and AGN activity reached their peak \citep{annurev:/content/journals/10.1146/annurev-astro-081811-125615}. This period marks a transitional phase in the evolution of the intergalactic medium (IGM), when the energy released by AGNs and star formation influences the transition from a predominantly neutral to a highly ionized IGM. Clusters and protoclusters discovered at this epoch are the precursors to today's massive galaxy clusters \citep{2016A&ARv..24...14O, 2019Sci...366...97U}, and they provide unique laboratories for studying AGN feedback and its large-scale effects \citep{Muldrew2015MNRAS.452.2528M}. These early-stage clusters typically host multiple AGNs and quasars, offering insights into how such activity shapes the large-scale evolution of galaxy ecosystems.

Lyman-alpha (Ly$\alpha$) emission serves as a powerful probe in these studies. Originating from the n=2 to n=1 transition in hydrogen, this emission line is the brightest in the ultraviolet spectrum of young, star-forming galaxies and AGNs \citep{2019SAAS...46.....V}. At high redshifts ($z>2$), where the majority of the IGM remains neutral, Ly$\alpha$ is the most prominent emission line, making it an essential tracer of hydrogen gas and feedback processes in the early Universe \citep{Steidel2011DiffuseGalaxies,2014Natur.506...63C}.

Recent advancements in wide-field spectroscopy have revolutionized our ability to study extended Ly$\alpha$ emission. Integral field units (IFUs) such as the Keck Cosmic Web Imager \citep[KCWI;][]{2018ApJ...864...93M}, the Multi Unit Spectroscopic Explorer \citep[MUSE;][]{2010SPIE.7735E..08B}, and the future BlueMUSE \citep{2019arXiv190601657R} simultaneously capture spectral and spatial information over a wide field of view (FoV), making them particularly well suited for detecting diffuse, low-surface-brightness (SB) emission that would be challenging to identify with slit spectroscopy or narrowband filters. These IFUs have revealed that extended Ly$\alpha$ halos — spanning from tens to hundreds of kiloparsecs — are ubiquitous in systems ranging from individual Ly$\alpha$-emitter (LAE) galaxies \citep[][]{2016A&A...587A..98W,2017MNRAS.465.3803V,2017A&A...608A...8L,2020A&A...635A..82L} to galaxy overdensities \citep{2022ApJ...926L..21D,2024A&A...683A..64A, 2024arXiv240801598B} and quasars \citep{2016ApJ...831...39B,2021MNRAS.502..494M}, suggesting a significant reservoir of cool gas in different environments. Moreover, the large-scale distribution of Ly$\alpha$ emission provides valuable insights into the physical conditions and properties of the cosmic web as a whole \citep{2018MNRAS.475.3854G,2018Natur.562..229W,2019NatAs...3..822M,2021MNRAS.504...16G,2023NatAs...7.1390M}. 

By examining Ly$\alpha$ emission in the context of quasar activity, particularly within the dynamic environments of high-redshift clusters and protoclusters, we gain a more comprehensive understanding of how AGNs influence their surroundings on both small and large scales. The combination of high sensitivity, high spectral resolution, and a large FoV allows IFUs to probe diffuse emission at very low SB levels, revealing how AGN feedback can shape the CGM and regulate gas accretion onto galaxies \citep{2017ApJ...837...71C,2019Sci...366...97U,2019MNRAS.482.3162A}. Moreover, combining IFU observations with complementary multiwavelength data — including radio interferometry for detecting jet-driven outflows, X-ray observations for tracing hot gas and AGN-driven shocks, and infrared imaging for obscured star formation and dust emission — provides a more complete picture of the interactions between RLAGNs and their environments. This multiwavelength approach enhances our ability to disentangle the effects of AGN-driven outflows from other processes shaping the CGM and IGM \citep{2023ApJ...958L..36C, 2023A&A...680A..70W}.

In this paper we focus on the high-redshift cluster CARLA J1017+6116 at $z = 2.8$ \citep{2018ApJ...859...38N}, which is part of the Clusters Around Radio-Loud AGN (CARLA) Survey \citep{2013ApJ...769...79W, 2014ApJ...786...17W}. We explore the complex interplay between the central quasar, its extended Ly$\alpha$ halo, and the broader cluster environment. Our study primarily relies on KCWI observations to investigate the spatially extended Ly$\alpha$ emission within the cluster, allowing us to analyze the kinematics and morphology of the cool gas in this overdense region. Additionally, we incorporate multiwavelength data — including \textit{Hubble} Space Telescope (HST) imaging, very long baseline interferometry (VLBI), and \textit{Spitzer}/Infrared Array Camera (IRAC) observations — to provide a comprehensive characterization of the cluster and the RLAGN. By examining the interactions between the quasar’s outflows, the CGM, and the surrounding cluster, we aim to understand how these components influence one another, particularly in shaping the evolution of galaxies within this dense environment. This integrated approach enables us to assess both the quasar’s feedback mechanisms and the large-scale structure of the cluster, offering new insights into the physical processes at work during this key epoch in cosmic history.

This paper is organized as follows: Section~\ref{sec:data} describes the survey, target selection, observations, and ancillary data. Section~\ref{sec:method} covers KCWI data reduction, the extraction of halo emission, the moment analysis of the Ly$\alpha$ halo, and the reanalysis of the VLBI data. Section~\ref{sec:results} presents our findings, including the morphology and kinematics of the Ly$\alpha$ halo and the results of the VLBI reanalysis. Section~\ref{sec:discussion} explores the broader implications of our results, particularly regarding quasar feedback mechanisms and the evolution of the cluster environment. Finally, Sect.~\ref{sec:conclusions} summarizes our key findings.

Throughout the paper we assume a flat $\Lambda$ cold dark matter cosmology with $\mathrm{H_0}=69.6\,\mathrm{km\,s^{-1}Mpc^{-1}}$, $\mathrm{\Omega_m}=0.286$, $\mathrm{\Omega_\Lambda}=0.714$ \citep{Bennett2014}, and $\mathrm{\Omega_b}\,h^{2}=0.02223$ \citep{Bennett2013}.

\section{The data}\label{sec:data}

\subsection{CARLA J1017+6116}\label{sec:data-carla}

CARLA J1017+6116 was spectroscopically confirmed as a cluster by \citet{2018ApJ...859...38N}. CARLA targeted 420 fields around RLAGNs using the \textit{Spitzer} Space Telescope to identify galaxy clusters through overdensities of galaxies with color IRAC1 (3.6\,$\mu$m) - IRAC2 (4.5\,$\mu$m) > -0.1\,mag. This selection efficiently identifies galaxies at $z > 1.3$ with $\sim$~ 95\% completeness and purity \citep{2013ApJ...769...79W, 2014ApJ...786...17W}. Approximately half of the CARLA fields display galaxy overdensities greater than $2\sigma$ above the field average. Sixteen of the most significant overdensities at $1.34 \leq z \leq 2.8$ were spectroscopically confirmed as probable clusters and highly probable clusters \citep{2016ApJ...830...90N, 2018ApJ...859...38N}. Among these, CARLA J1017+6116, at $z = 2.801$ \citep{2018ApJ...859...38N}, is the highest redshift spectroscopically confirmed CARLA highly probable cluster.

\begin{figure}
\includegraphics[trim={2cm 1cm 2cm 2cm},clip,width=\columnwidth]{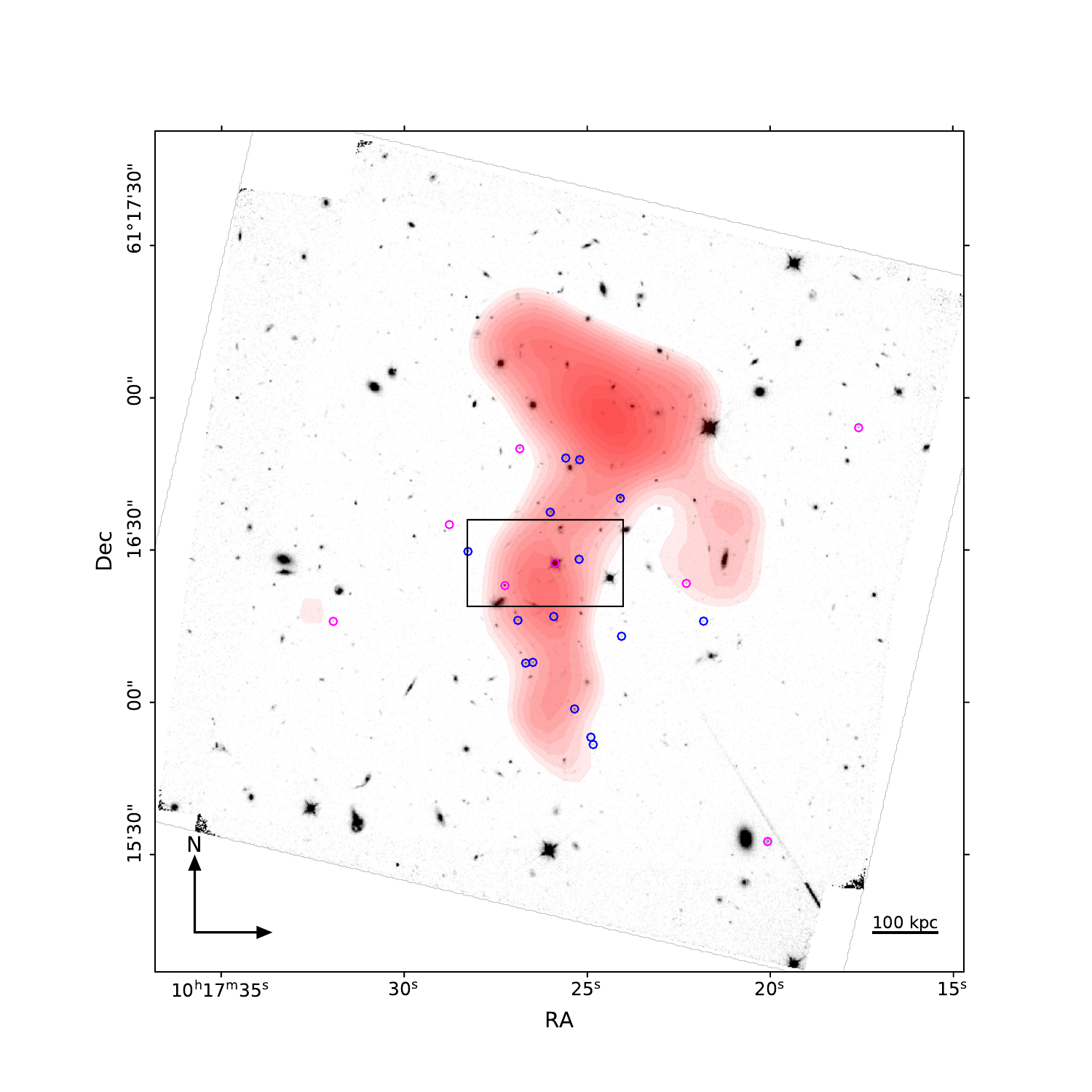}
\caption{HST F140W image, centered on CARLA J1017+6116. 
We show as circles the galaxies that are either spectroscopically confirmed at the cluster redshift \citep[blue circles;][]{2018ApJ...859...38N} or selected as statistically belonging to the clusters \citep[magenta circles;][]{2023A&A...670A..58M}. The black rectangle indicates the KCWI FoV. Red contours represent galaxy overdensity starting from ${\rm S/N =3,}$ as indicated in \citet{2023A&A...670A..58M}. The higher overdensity peak north of CARLA J1017+6116 is associated with a galaxy overdensity at z=1.235 discovered by \citet{2018ApJ...859...38N}, while the central overdensity is associated with the CARLA cluster and our KCWI observations. The north overdensity could not be excluded in the image contours due to the specific color cuts implemented to identify overdensities.}
\label{fig:hst}
\end{figure}

\citet{2023A&A...670A..58M} analyzed the cluster galaxy overdensity and morphology. They constructed an overdensity map of galaxies with  \textit{Spitzer} IRAC1 and IRAC2 channels $(\text{IRAC1} - \text{IRAC2}) > -0.1$ and $\text{IRAC1} < 22.6$\,mag, and selected galaxies in over-dense regions with limited foreground and background contamination. This process effectively identifies about 90\% of galaxies at $z > 1.3$, with only 10\% contamination from non-cluster members. 
They defined the galaxy projected overdensity as $\Sigma$~=~$\frac{N_{gal}-N_{bkg}}{N_{bkg}}$ and  calculated the overdensity signal-to-noise ratio (S/N) as ${\rm S/N}$~$\equiv \frac{N_{gal}-N_{bkg}}{\sigma_{N_{bkg}}}$. Figure \ref{fig:hst} shows their S/N map, starting from ${\rm S/N =3}$
This map reveals significant galaxy overdensities, with the highest$\text{ S/N}$ detected at approximately 0.60\arcmin\ north of the quasar. This corresponds to a prominent overdensity spectroscopically confirmed at $z = 1.235$ \citep{2018ApJ...859...38N}. 
The secondary peak, which this paper focus on, is close to the AGN and shows a $\text{S/N} = 14$ galaxy overdensity, which has been classified as a highly probable galaxy cluster with a total stellar mass of $M_*=10^{12.6} M_\odot$, and an estimate on the total halo mass of $M_h=10^{14.6} M_\odot$ \citep{2023A&A...670A..58M}, which is derived from models and might be an upper limit. 
CARLA J1017+6116 presents a significant population of early-type galaxies (ETGs) and mergers 
and the correlation between its ETG fractions and galaxy density is consistent with a morphology-density relation already in place at $z \approx 3$, though further studies are required to confirm its universality at such high redshifts \citep{2023A&A...670A..58M}.

\subsection{CARLA observations}

CARLA J1017+6116 was observed with the \textit{Spitzer} Space Telescope's IRAC on May 23, 2012, using channels 1 and 2 ($3.6\,\rm{\mu m}$ and $8\,\rm{\mu m}$, respectively) under Proposal Cycle 11 (Program ID 80254; P.I. D. Stern). The total exposure times were 1000\,s in IRAC1 and 2100\,s in IRAC2, providing a similar depth in the two channels. The 95\% detection completeness limit was achieved at $\rm{IRAC1 = 22.6\,mag}$ and $\rm{IRAC2 = 22.9\,mag}$. Regions with limited coverage ($< 85\%$) were excluded from our analysis. The point spread function (PSF) is 1.95\arcsec\ and 2.02\arcsec\ in IRAC1 and IRAC2, respectively. Full details of the image processing can be found in \citet{2013ApJ...769...79W, 2014ApJ...786...17W}.

The cluster was followed up with the HST to obtain Wide field camera 3 (WFC3) HST F140W ($H_{140}$) images and {\it G141} grism spectroscopy (Proposal ID 13740; P.I. D. Stern). The total exposure time was 1000\,s for the $H_{140}$ images and 4000\,s for the \textit{G141} grism spectroscopy. The details of the HST data reduction are provided in \citet{2016ApJ...830...90N}.

For this paper analysis, we observed the cluster with KCWI (P.I. Chris Martin) on the Keck II 10-meter telescope at Maunakea, Hawaii. The observations were performed on November 21, 2017, with a total exposure time of 50 minutes. The data were captured using a binning configuration of $2\times2$, with a CCD gain set to 0.145\,e-/DN. The airmass during the observations was 1.47, and the full width at half maximum (FWHM) of the guider was measured at $\sim 1.044\arcsec$, ensuring a high-quality spatial resolution across the field.
The observations were conducted using the blue medium grating on KCWI, which covers a spectral range of approximately 3500\,\AA\ to 5500\,\AA\ in the medium-resolution mode, with a wavelength width per pixel of 0.5\,\AA. In this configuration, the instrument's FoV was $20\arcsec\times 33\arcsec$ (about $160\times 264$ in pkpc at $z=2.8$). Data reduction involved standard procedures such as bias subtraction, overscan correction, flat-fielding, and flux calibration. Since the detection of extended low-SB emission can be sensitive to noise amplification in smoothed maps, we carefully assessed the significance of the detected features. Potential noise artifacts were mitigated by implementing strict signal-to-noise thresholds and visually inspecting the maps for spurious detections. Further details on the noise treatment and robustness tests are provided in Sect. \ref{sec:reduction}.

\subsection{Ancillary data}\label{sec:data-achillary}

\begin{figure}
\includegraphics[trim={.4cm 0cm .1cm 0cm},clip,width=\columnwidth]{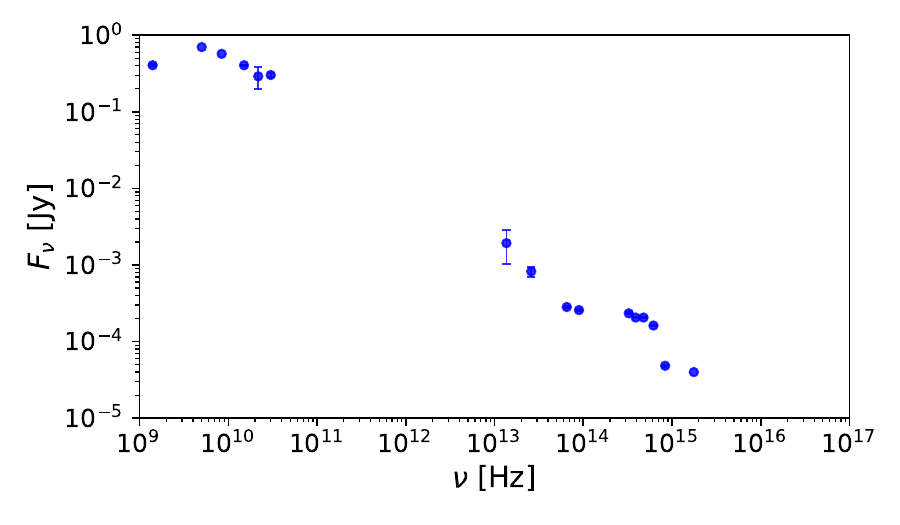}
\caption{Spectral energy distribution of the J1017+6116 AGN, spanning radio to X-ray frequencies.  The radio variability is likely driven by shock waves in the jet and Doppler boosting effects, which may correlate to episodic energy injections or changes in viewing angle (see Sect. \ref{sec:qso-feedback}).
}
\label{fig:qso_sed}
\end{figure}

\begin{figure}
\includegraphics[trim={.5cm 0cm 2.2cm 0cm},clip,width=\columnwidth]{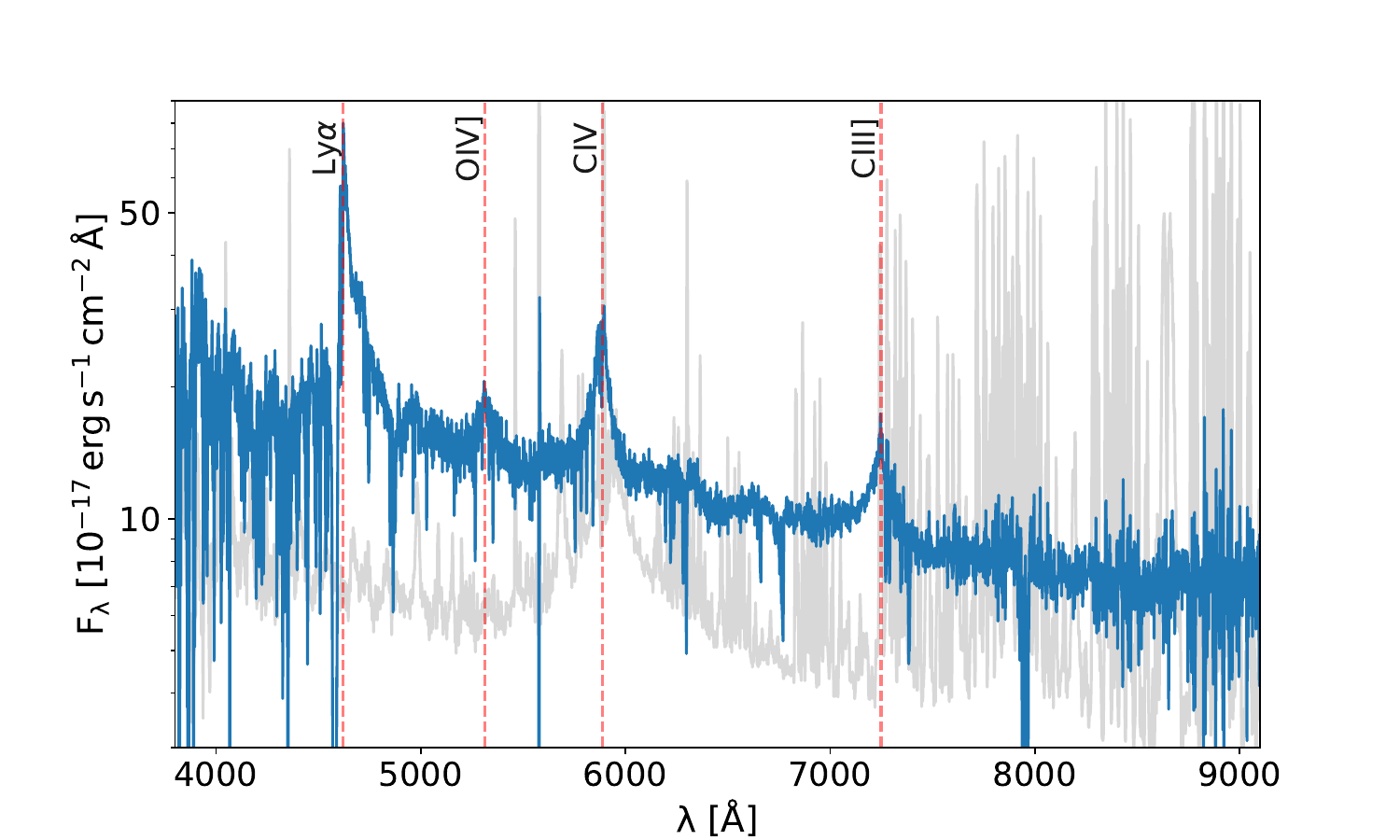}
\caption{SDSS spectrum of the J1017+6116 AGN. The key emission lines are indicated with dashed red lines for reference. The background sky flux is shown in gray.}
\label{fig:qsospec}
\end{figure}

\begin{figure}
\includegraphics[trim={0cm 0cm 0cm 0cm},clip,width=\columnwidth]{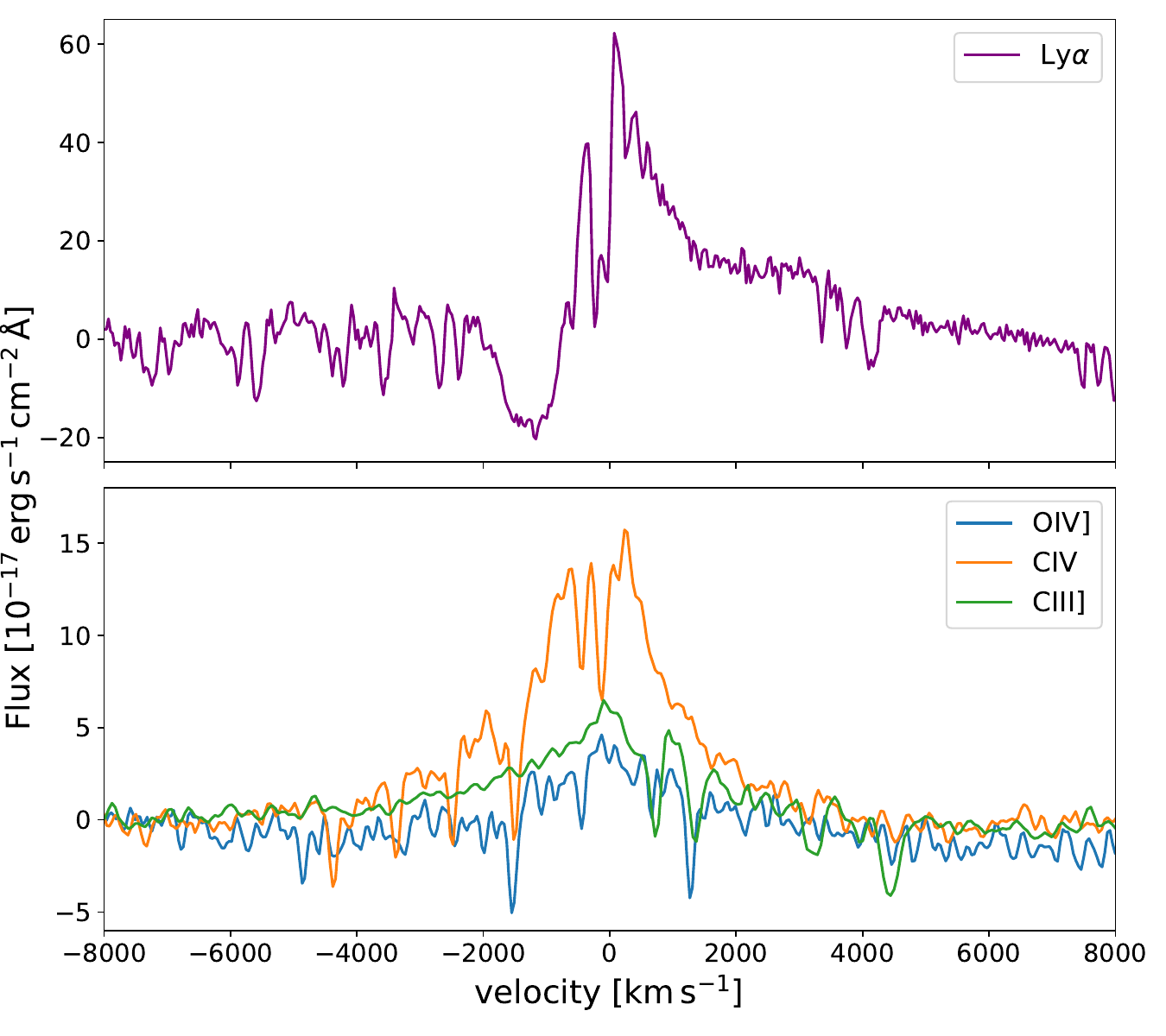}
\caption{SDSS continuum-subtracted spectra of the J1017+6116 AGN Ly$\alpha$ (upper panel), OIV], CIV, and CIII] lines (bottom panel).}
\label{fig:QSO_lines_spec}
\end{figure}

Extensive ancillary observations around the quasar J1017+6116 span a wide range of frequencies, from X-ray to radio wavelengths, as illustrated in Fig. \ref{fig:qso_sed}. These datasets offer a comprehensive understanding of the quasar's spectral characteristics. Key observations include:

\begin{itemize}
    \item Sloan digital sky survey (SDSS) optical spectra with key emission lines marked, as shown in Figs. \ref{fig:qsospec} and \ref{fig:QSO_lines_spec} \citep{2000AJ....120.1579Y}. 
    \item VLBI interferometry data across multiple epochs in S, X, and C bands (2, 5, and 8 GHz respectively) obtained from the Radio Fundamental Catalog
\citep[RFC;][]{2024arXiv241011794P}, detailed in Fig. \ref{fig:VLBI}.

\item Very large array (VLA) observations include data at 74 MHz from the VLA low frequency sky survey (VLSS) \citep{2007AJ....134.1245C}, with $80''$ resolution and $\sim$0.1 Jy beam$^{-1}$ rms; at 1.4 GHz from NRAO VLA sky survey (NVSS) \citep{1998AJ....115.1693C}, with $45''$ resolution and $\sim$0.45 mJy beam$^{-1}$ rms; and at 8.4 GHz from the Combined radio all-sky targeted eight-GHz survey (CRATES) \citep{2007ApJS..171...61H}, which offers subarcsecond structure and spectral indices for over 11,000 extragalactic sources.

\item Low-frequency array (LOFAR) observations from the Two-metre sky survey (LoTSS) Data Release 2 \citep[DR2;][]{2022A&A...659A...1S}.

\item Mid-infrared data from the Wide-field infrared survey explorer (WISE), as detailed in Sect. \ref{sec:data-carla} \citep{2010AJ....140.1868W}.

\item Optical data in the \(ugriz\) bands from SDSS \citep{1996AJ....111.1748F}.

\item X-ray observations from ROSAT 
(0.1 to 2.4 keV), as part of the ROSAT all-sky survey bright source catalog (RASS) \citep{2007A&A...476..759B}, detect a very low X-ray flux from our quasar, resulting in a bolometric luminosity of $3.65 \times 10^{12} \pm 1.57 \times 10^{12} \, L_\odot$. This low X-ray luminosity suggests either that the quasar is  intrinsically X-ray weak or that the X-ray emission is heavily obscured. The low X-ray emission also constrains the potential contribution of inverse Compton scattering as a dominant mechanism, implying that beaming effects or large-scale jet structures are not as influential in this system, or that the jet is misaligned relative to our line of sight (LoS).

\end{itemize}

\subsection{Previous studies on the quasar J1017+6116}
In addition to the research on CARLA J1017+6116 properties by \citet{2023A&A...670A..58M} and \citet{2018ApJ...859...38N}, several studies have extensively analyzed its quasar:

\begin{itemize}
    \item The quasar is part of the study by \citet{2022MNRAS.511.5436D}, where it is categorized as a flat spectrum radio quasar.
    \item \citet{2017A&A...597A..79P} provide detailed measurements for the FWHM of C\,III (4403.5\,km/s) and C\,IV (3325.2\,km/s), along with the absorption index of the C\,IV trough (AI C\,IV) at 106.6\,km/s.
    \item Shen et al. (2011) summarized data for 105,783 quasars from SDSS\,(DR7), including continuum and emission line measurements across critical spectral regions like H$\alpha$, H$\beta$, Mg\,II, and C\,IV. For J1017+6116, a black hole mass of \(9.65\pm0.04\ \log_{10}\mathrm{M}_\odot\) was estimated using the C\,IV-based mass estimator.
    \item \citet{2011ApJS..194...45S} noted the bolometric luminosity of J1017+6116 suggests mass outflow rates up to \(10^4\ \text{M}_\odot\,yr^{-1}\), typical for hyper-luminous quasars at \(z=2-3\) exhibiting ionized winds on the order of \(10^3\ \mathrm{M_\odot\,yr}^{-1}\).
    \item The study by \citet{2022ApJS..260....4P} developed a method for estimating parsec-scale jet directions using a model-fitting technique on calibrated VLBI visibilities. Applying this to J1017+6116, the estimated position angles (PAs) at 2, 5, and 8\,GHz are \(130\pm4\), \(-108\pm5\), and \(-104\pm3,\) respectively, with an averaged mean PA of \(-136\pm3\).
    \item \citet{2017MNRAS.472.1850G} discovered a damped Ly$\alpha$ absorber (DLA) in the LoS to J1017+6116 at $z=2.7675$ with a log column density (\(\log_{10}(\text{N}_{\text{HI}})\)) of 20.4422.
\end{itemize}

\noindent A summary of the available information on J1017+6116 is provided in Table \ref{tab:qso}.

\begin{table}
        \centering
        \caption{Properties of the quasar J1017+6116.}
        \label{tab:qso}
        \begin{tabular}{ccc} 
                \hline
               RA & (J2000) & 10:17:25.887 \\
               Dec & (J2000) & +61:16:27.496 \\
               $z_\mathrm{spec}$ &   & 2.7681{\color{red} } \\
               $\mathrm{M_{BH}}$ &  [$\mathrm{log_{10}\, M_\odot}$] & $9.65\pm0.04$ \\
               SFR & [$\mathrm{M_\odot\,yr^{-1}}$] & 500 {\color{red} } \\
               CIV FWHM & [$\mathrm{km\,s^{-1}}$] & 3325.2 {\color{red} } \\
               CIII FWHM & [$\mathrm{km\,s^{-1}}$] & 4403.5 {\color{red} } \\
               $\mathrm{L_{bol}}$ & [$\mathrm{log_{10}\,erg\,s^{-1}}$] & $47.400\pm0.005$ \\
               Eddington ratio &  & $0.15\pm0.32$ \\
            Jet direction & degrees & $-136\pm3$ \\
                \hline
        \end{tabular}
\end{table}

\section{Methodology}\label{sec:method}

In this section we describe the methodology adopted for our analysis, focusing on the data reduction and the extraction of extended emission centered on the quasar, using a smoothing technique to enhance faint, diffuse structures in the data effectively. We also focus on the extraction of spectral line moments, specifically the velocity offset and FWHM, to analyze the kinematic behavior of the gas. Additionally, we perform a reanalysis of the VLBI data to investigate the nature of the radio emission, assessing the presence of jet structures and variability.

\subsection{KCWI data reduction}\label{sec:reduction}

We performed the initial data reduction using the standard KCWI Data Reduction Pipeline\footnote{\href{https://github.com/kcwidev/kderp}{https://github.com/kcwidev/kderp}}, which converts raw 2D science frames into flux-calibrated, 3D cubes with real-world coordinate systems and wavelengths. The pipeline includes bias and overscan subtraction, gain correction, trimming, and cosmic-ray removal. Dark subtraction and scattered light removal were also performed. Calibration images were used to define the geometric transformations that map each pixel in the 2D image to slice, position, and wavelength coordinates. Flat-field and illumination corrections were applied, and a spectrophotometric standard star observation was used to convert detector counts to physical flux units. This process yielded 3D, flux-calibrated data cubes for each individual exposure.

For cube correction and co-adding, we employed \texttt{CWITools} \citep{2020arXiv201105444O}, a Python toolkit specifically designed for KCWI data analysis. Individual exposure cubes were first corrected by adjusting their world-coordinate system and trimming them. The RA, Dec coordinate system for each frame was corrected using the known location of a visible source in the field. The wavelength axis was refined using the positions of known sky emission lines. The cubes were then trimmed to retain only the wavelength range shared by all slices, and edge pixels were removed from the spatial axes.

\texttt{CWITools} co-added the corrected and cropped input cubes by computing the footprint of each input pixel on the co-add frame and distributing flux accordingly. The on-sky footprint of each input frame was determined, and a new world-coordinate system representing the co-add frame was constructed to encompass all input data with a 1:1 aspect ratio. The wavelength axes of the input cubes were aligned using linear interpolation, allowing the co-adding process to become 2D. This method ensures that the input frames — regardless of their spatial resolution or PA — are accurately combined, preserving data integrity and allowing for variance estimation.

To mitigate the influence of bright quasar emission on the detection of extended features, we subtracted the PSF of the quasar spectrum and masked a $3\arcsec$ diameter region centered on the quasar to eliminate residual flux. This calibration and data reduction approach ensured high-quality and precise data for our analysis.

\subsection{Extraction of extended emission}

The extraction of the extended emission was performed using the adaptive kernel smoothing (AKS) technique \citep[e.g.,][]{2019NatAs...3..822M,2020ApJ...894....3O,2021A&A...649A..78D}. This method enhances faint, diffuse structures with a low S/N while preserving the integrity of spatially resolved features. AKS adaptively adjusts the smoothing scale by applying Gaussian smoothing across the datacube, effectively highlighting extended emission while minimizing the risk of spurious noise amplification.

The AKS method begins by smoothing the data with a Gaussian kernel at a predefined minimum scale. Pixels that meet a specified S/N are detected at this scale. The smoothed values of these detected pixels are then subtracted from the working image and added to a detection image. To prevent re-detection in subsequent iterations, detected pixels are masked. The smoothing kernel size is increased iteratively until all pixels are masked or a maximum kernel size is reached. The final detection image combines signals detected at various scales, ensuring that bright emission regions are minimally smoothed while background areas with little or no signal are smoothed at larger scales.

For 3D, such as a datacube, the AKS process can be applied to both spatial and spectral dimensions. However, to optimize the extraction of the Ly$\alpha$ halo while minimizing spectral contamination, we chose to apply AKS only to the spatial dimensions. This approach preserves the spectral resolution while enhancing faint, extended emission. The smoothing window varied from $0.3\arcsec$ (no smoothing) to $1.5\arcsec$, with a S/N of 6. The convolution was applied incrementally, starting from the original data and progressively increasing the smoothing radius up to 5 pixels.

We estimated the variance of the AKS-detected emission by convolving the variance of the datacube with the square of the kernel on each layer. While this method assumes uncorrelated noise between voxels — an assumption that does not fully apply to KCWI datacubes — we rescaled the variance based on the original unsmoothed data to better reflect correlated noise structures.

To ensure that the AKS method did not introduce artificial structures or amplify noise peaks, we conducted a series of reliability tests. First, we performed noise robustness checks by comparing the smoothed emission maps to their unsmoothed counterparts, identifying any discrepancies that could arise from excessive noise amplification. Regions with anomalously low FWHM values or extreme velocity offsets (e.g., $\pm500$ km/s pixels) were flagged and reevaluated. Additionally, we applied the AKS technique to spectral slices offset from the Ly$\alpha$ emission, serving as a randomized control to confirm that no similar extended structures appeared in noise-dominated spectral regions. Furthermore, we tested different smoothing kernels and threshold levels to verify that the detected features were not highly sensitive to the specific AKS parameters used. These steps collectively ensure that the extended Ly$\alpha$ emission features reported in this study are robust and not artifacts of noise amplification or systematic biases in the smoothing process.

To ensure sufficient spectral resolution and minimize noise from low-information pixels, we applied a voxel-count threshold to the AKS segmentation mask. Only pixels with at least five connected voxels were retained in the final moment maps. This threshold was chosen to balance spatial completeness and spectral reliability across the extracted Ly$\alpha$ halo.

\subsection{Halo spectral line moments}\label{sec:moments}
The extraction of spectral line moments from the datacubes processed through AKS focuses on the first two moments: the velocity offset with respect to the flux-weighted mean of the line emission (centroid) and the FWHM. The centroid measure provides the velocity offset of each spaxel with respect to the flux-weighted centroid position of the spectral line in the spectrum, offering insights into the relative velocity of the emitting gas. The FWHM, on the other hand, quantifies the width of the spectral line at half of its maximum intensity. This measure is crucial for assessing the velocity dispersion within the emitting material. A broader FWHM indicates a greater spread in the velocities of the gas, which can be indicative of various dynamic processes occurring within the cluster, such as turbulent motions, gravitational interactions, or quasar feedback.

\subsection{Reanalysis of VLBI data}\label{sec:vlbi-analysis}

The original VLBI analysis of J1017+6116 by \citet{2022ApJS..260....4P}, mentioned in Sect. \ref{sec:data-achillary}, was part of a broader statistical study characterizing the general radio properties of quasars. However, it did not provide a detailed, source-specific analysis and assumed that the radio morphology was linked to jet emission. Their method estimated the parsec-scale jet PA by fitting a two-Gaussian model to VLBI visibilities, identifying the brightest component as the core and the secondary as the jet feature. The PA was measured as the angle between them, with intensity-based core selection when uncertainties arose. They then computed a simple average of PA values across multiple epochs for each frequency before taking the final PA as the average across all available frequencies.

When examining their methodology and the PA values for J1017+6116, we found inconsistencies across frequencies, raising concerns about whether the standard approach accurately reflected the quasar’s behavior. This motivated a reanalysis to verify whether the detected PA was real or an artifact of the method. For this purpose, we used pre-reduced VLBI data at multiple frequencies (S band at 2 GHz, C band at 5 GHz, and X band at 8 GHz), observed across different epochs. We examined the VLBI data on arcsecond scales but did not identify a clear jet structure or any significant emission beyond the central source in any of the available observations. Therefore, we extracted $80\times80$ milliarcsecond (mas) cutouts centered on the quasar, ensuring the primary emission region was fully enclosed while avoiding contamination from unrelated background sources. This approach allowed us to focus on the quasar and its immediate surroundings. 

The flux density for each cutout was determined by integrating the emission above a 3-sigma threshold in a contiguous region centered on the quasar’s peak emission. The boundary of this region was defined by the 3-sigma threshold based on the background noise level in each cutout. This method provided a reliable measurement of the quasar’s primary radio emission while minimizing contamination from noise or unrelated sources. Background noise levels were cross-checked across different frequencies and epochs to account for variations in observation sensitivity.

In addition to flux measurements, we analyzed the PA of the radio emission using two methods: (i) fitting a two-Gaussian model to the source, following the approach of the original study, and (ii) fitting a single Gaussian fixed at the main source (centered in all our data) and subtracting it from the measured flux density described above. This approach enabled us to track both emission and PA variations over time and across different frequencies.

\section{Results}\label{sec:results}

\begin{figure*}
\includegraphics[trim={3cm 0.5cm 2.5cm 0cm},clip,width=2\columnwidth]{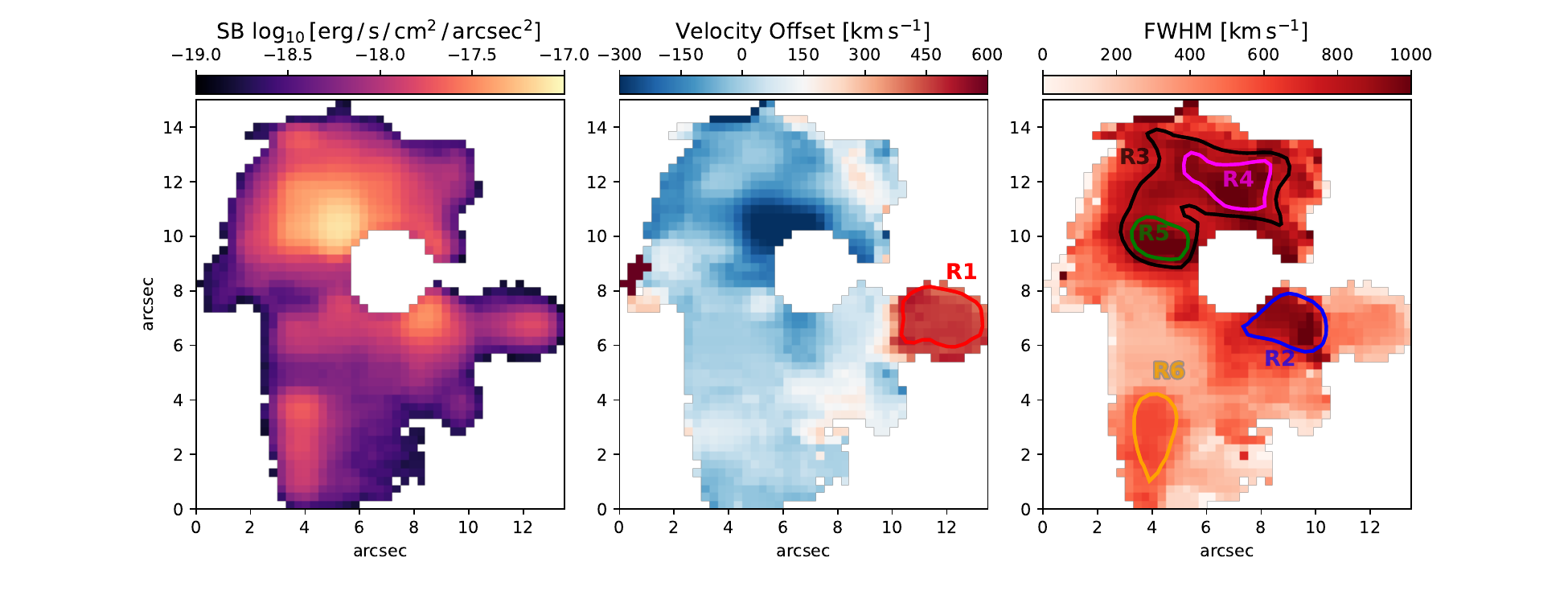}
\caption{
LAE halo detected in the KCWI datacube. The panels correspond to the first three moment maps of the extended emission: SB, flux-weighted centroid with respect to the mean velocity, and the FWHM of the spectral line. The emission was extracted from the datacube using the AKS technique, which is specifically
designed to separate extended emission with low SB from the background noise. Region R1, identified based on the velocity offset map (center panel), corresponds to a redshifted structure with a velocity offset exceeding 400 km s$^{-1}$. Regions R2 to R6, defined from the FWHM map (right panel), correspond to areas with broad Ly$\alpha$ emission: R2 includes FWHM values above 700 km s$^{-1}$, R3 above 800 km s$^{-1}$, R4 and R5 above 900 km s$^{-1}$, and R6 above 500 km s$^{-1}$. A supplementary figure in the appendix provides an unsmoothed visualization of the moments (Fig. \ref{fig:original_moments}), along with the S/N and the number of voxels per pixel within the segmentation mask for reference (Fig. \ref{fig:SNR_Nvox}).
}
\label{fig:moments} 
\end{figure*}

\begin{figure}
\raggedright
\includegraphics[trim={1.7cm 1cm 1cm 1cm},clip,width=1.05\columnwidth]{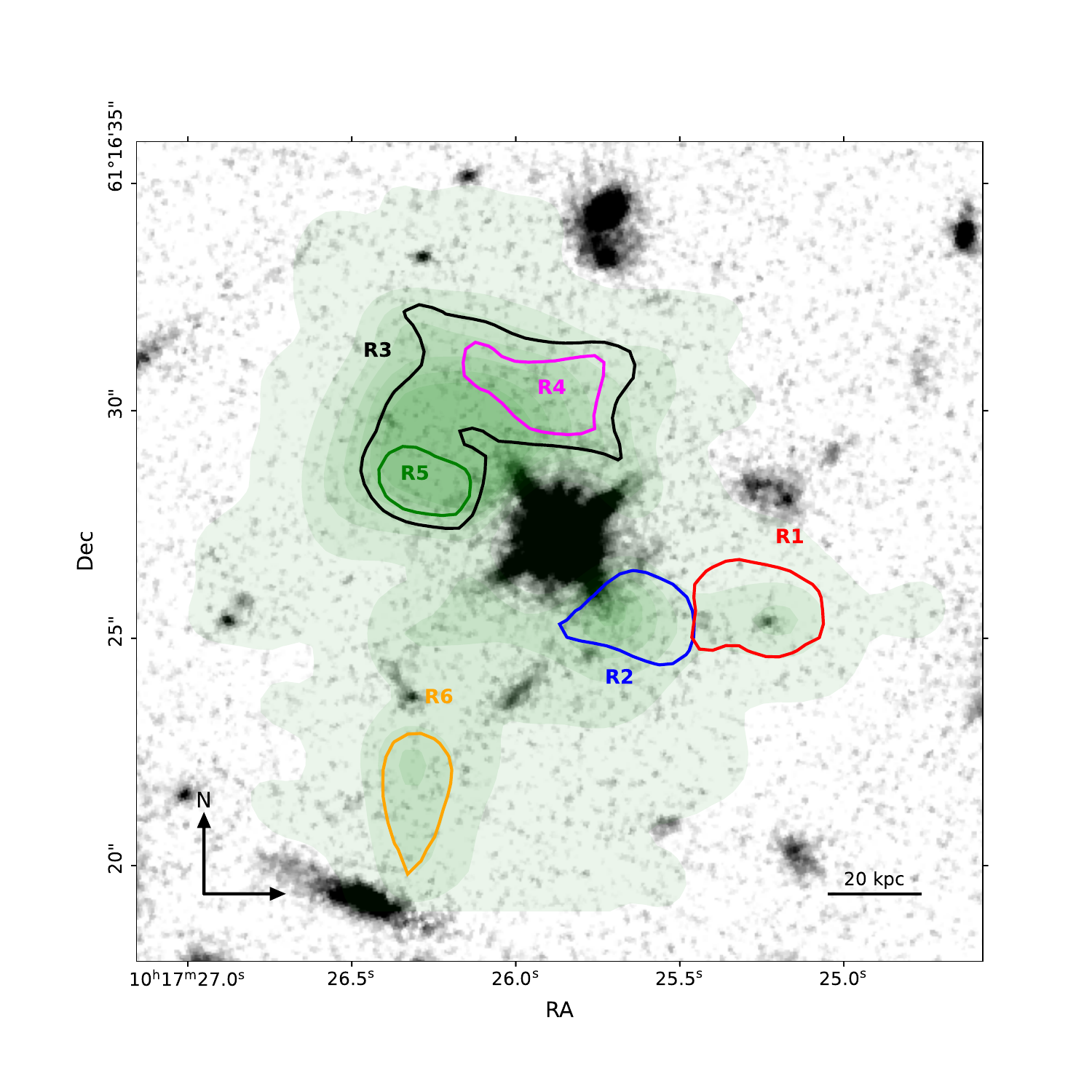}
\caption{HST F140W image of the area surrounding the Ly$\alpha$ halo, with SB green contours representing Ly$\alpha$ emission levels overlaid. The contours, smoothed using a Gaussian filter of 3 pixels, highlight the extent of the emission. We defined six distinct regions (R1 to R6) based on their kinematic and morphological properties and applying the same Gaussian smoothing. Region R1 (red), characterized by a high velocity offset (exceeding 400 km s$^{-1}$), is interpreted as an independent halo associated with a continuum source. Regions R2 to R6 are defined by contours of the FWHM image, with R2 (blue) corresponding to FWHM values above 700 km s$^{-1}$, R3 (black) above 800 km s$^{-1}$, R4 (magenta) and R5 (green) above 900 km s$^{-1}$, and R6 (gold) above 500 km s$^{-1}$
}
\label{fig:regs}
\end{figure}

\subsection{Ly$\alpha$ halo morphology}
Using the AKS technique, we extracted the extended Ly$\alpha$ halo surrounding the radio-loud quasar in CARLA J1017+6116 (Fig. \ref{fig:moments}, left panel). The Ly$\alpha$ halo displays pronounced morphological asymmetries and covers a substantial portion of the KCWI FoV. Evidence suggests that the halo extends beyond the observed limits, particularly in the southern direction, with a minimum extent of 16 arcsec to a SB level of $10^{-19}$ $\mathrm{erg\,s^{-1}\,cm^{-2}\,arcsec^{-2}}$, corresponding to approximately 128 physical kpc at the redshift of $z=2.8$. The total observed Ly$\alpha$ luminosity of the halo, without absorption correction\footnote{Reconstructing intrinsic Ly$\alpha$ emission is common in spectral analysis, using methods like Monte Carlo radiative transfer modeling to correct for scattering \citep{2021A&A...649A..78D} or Gaussian decomposition for flux estimation \citep{2023A&A...680A..70W}. Corrections typically range from 10\% to 90\%, depending on system conditions. In our case, the red-asymmetric profiles suggest scattering or outflow effects, but since our focus is on broader trends rather than precise flux reconstructions, a detailed correction will not impact our main conclusions.}, is ${43.35\pm0.05}\,\rm{log_{10}\,L_\odot}$. 

The SB profile reveals two primary regions of enhanced emission: R3 and R2 (Fig. \ref{fig:regs}), located northeast and southwest of the quasar, respectively, both showing significant morphological asymmetry. Region R3 is the brightest part of the halo, situated near the quasar, while R2 shows extended emission toward the southwest. The bright regions suggest the presence of localized interactions or external influences on the CGM. The asymmetry in these regions may be indicative of dynamic processes affecting the halo.

\subsection{Ly$\alpha$ halo kinematics}

\begin{figure}
\includegraphics[trim={0cm 0cm 0cm 0cm},clip,width=\columnwidth]{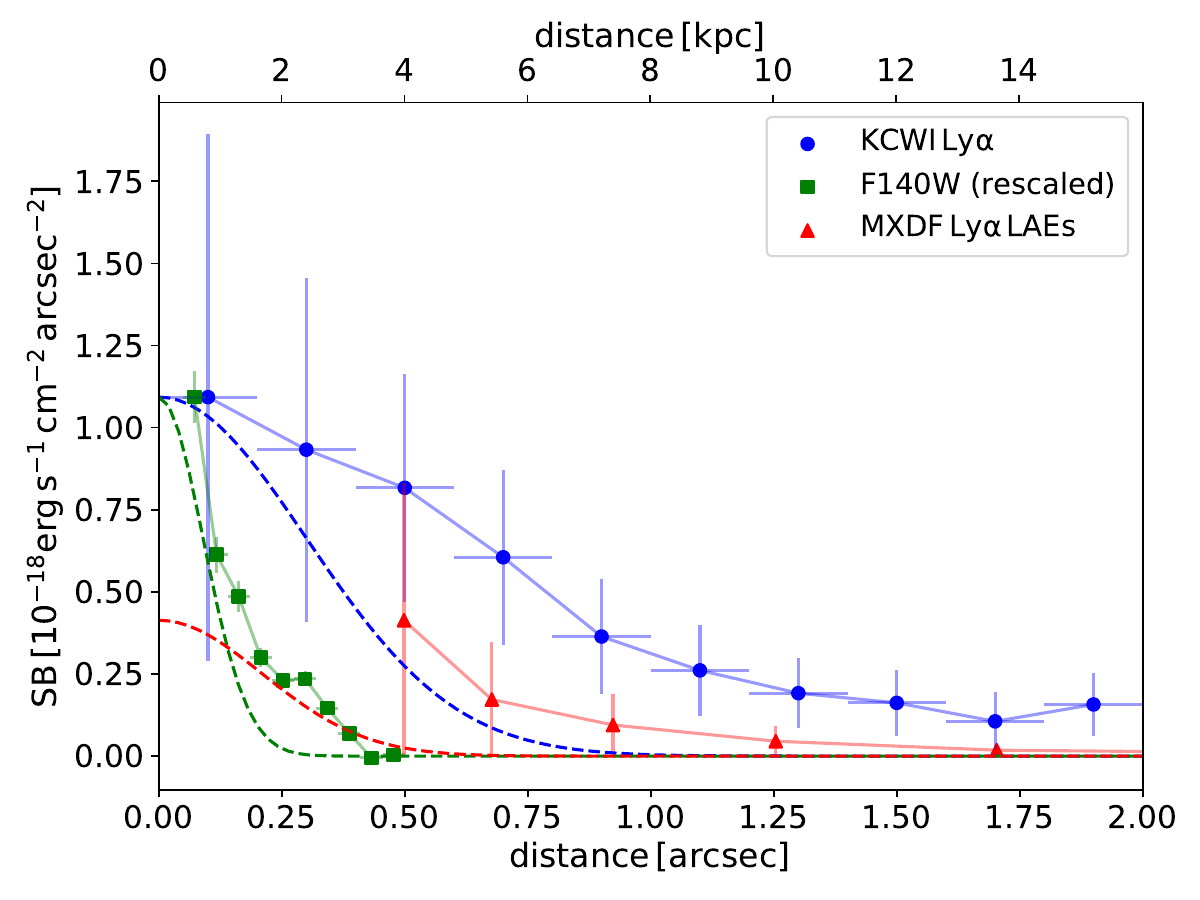}
\caption{
SB radial profile centered on the peak of the Ly$\alpha$ emission found to the southwest of the quasar, corresponding to a LAE in region R1 (see Fig. \ref{fig:regs}). The blue circles represent the radial profile extracted from the KCWI data.
The green squares correspond to the HST F140W profile, which has been rescaled to match the Ly$\alpha$ profile to aid visual comparisons and highlight the stellar component associated with the source. The red triangles show the mean SB profile of LAEs from the MXDF MUSE field \citep{2024A&A...688A..37G}, providing a comparison between the emission profile of region R1 and a typical LAE environment. This comparison emphasizes the enhanced SB of the Ly$\alpha$ halo in region R1 relative to the average LAE population, potentially suggesting additional contributions from the CGM or interactions with the quasar's environment.}
\label{fig:laeHST}
\end{figure}

The kinematic structure of the Ly$\alpha$ halo was examined through velocity offset and FWHM maps (Fig. \ref{fig:moments}). Most of the halo exhibits relatively quiescent kinematics, with little velocity shift relative to the quasar. However, a distinct redshifted region (R1) in the southeast part of the halo shows a velocity offset of approximately $450\,\rm{km\,s}^{-1}$ (Fig. \ref{fig:moments}, central panel). This region coincides spatially with a continuum source detected in the HST F140W imaging, confirming its nature as a LAE (Fig. \ref{fig:regs}).

Figure \ref{fig:laeHST} presents the SB radial profile of this LAE, extracted from KCWI data and compared to both the $H_{140}$ continuum profile and the average LAE profile from the MUSE extremely deep field (MXDF) survey. The LAE in region R1 displays a more extended and enhanced Ly$\alpha$ SB than typical LAEs, suggesting additional contributions from the CGM or interactions with the quasar’s environment. The spatial overlap of the Ly$\alpha$ emission and the continuum source in HST imaging supports the presence of an underlying stellar component, with the extended Ly$\alpha$ emission potentially tracing infalling gas from a filamentary structure or CGM interactions. The kinematics and brightness excess of this LAE relative to the average LAE population indicate that it may be dynamically influenced by the quasar’s feedback processes, reinforcing the idea that the CGM in this system is being actively shaped by the central AGN.

The FWHM map reveals significant broadening in certain regions of the halo, particularly near the quasar. Regions R4 and R5 show broad Ly$\alpha$ emission lines, with FWHM values exceeding $900\,\rm{km\,s}^{-1}$. These regions are dynamically distinct, likely indicating localized sources of gas acceleration, such as turbulence or feedback-driven motions. The absence of a coherent rotational pattern suggests that the halo is not undergoing steady rotation but is instead shaped by external influences, possibly inflows or outflows within the CGM.

The broad emission features may reflect complex interactions between the CGM and external processes, such as gas inflows from the surrounding environment or mechanical and radiative feedback from the RLAGN. However, without a clear spatial connection between the VLBI radio features and the halo kinematics, the precise origin of the broad Ly$\alpha$ emission remains uncertain. The observed velocity structure and asymmetries reinforce the idea that the halo is shaped by a combination of environmental interactions and AGN-driven feedback mechanisms.

\subsection{Spectral features of halo regions}

\begin{figure}
\includegraphics[trim={1cm 0cm 0.9cm 1cm},clip,width=1\columnwidth]{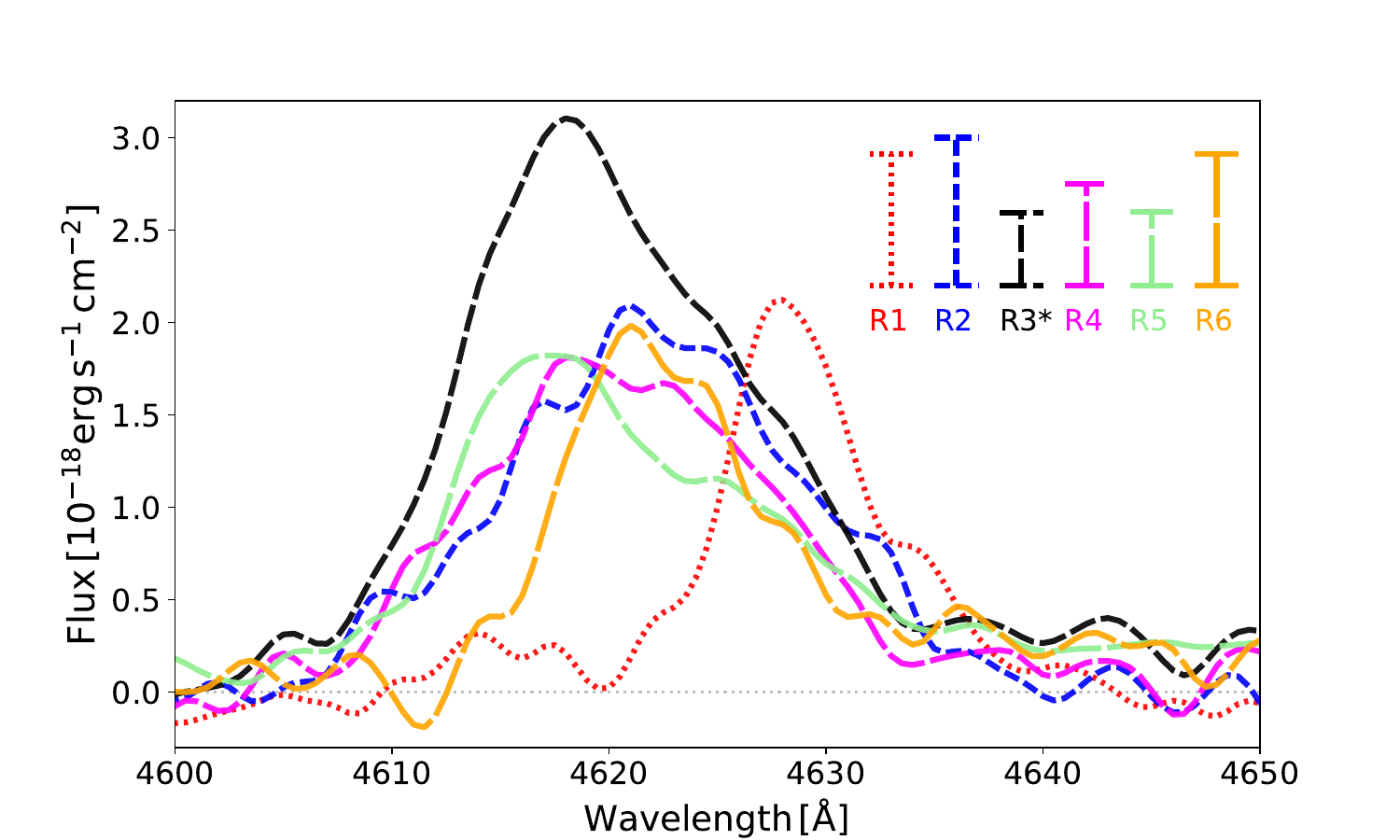}
\caption{
Ly$\alpha$ spectra for each of the regions in Fig. \ref{fig:regs}, showing flux as a function of wavelength. We normalized R3 by a factor of 3.5 to fit within the plot. The spectra were extracted from the original datacube (without AKS), but we applied a Gaussian smoothing of 2 pixels for visualization purposes. The error bars in the top-right corner represent 2$\sigma$ for each region, estimated from wavelength ranges excluding the emission line ($4600 - 4650$ \AA).
}
\label{fig:regspecs}
\end{figure}

Figure~\ref{fig:regspecs} presents the Ly$\alpha$ spectra extracted from each of the defined halo regions (R1 to R6; see Figs.~\ref{fig:moments} and~\ref{fig:regs}). Despite variations in morphology and kinematics across the halo, the overall spectral shape of Ly$\alpha$ emission remains consistent across all regions. Each region exhibits a slightly asymmetric profile with a redward skew. The peak emission wavelength varies by several hundred $\rm{km\,s^{-1}}$ depending on the region, reflecting the velocity offset structure shown in Fig.~\ref{fig:moments}.

Regions R4 and R5, where the Ly$\alpha$ FWHM exceeds 900 km s$^{-1}$, exhibit broader line profiles compared to the rest of the halo. The Ly$\alpha$ emission in these regions shows no strong secondary peaks or significant absorption features within the spectral resolution of KCWI.

Region R1, which corresponds to an independent LAE offset by approximately 450 km s$^{-1}$ from the systemic velocity of the quasar, presents a distinct profile compared to the rest of the halo. The spectral width in R1 is narrower than in regions R2-R6, and its peak emission is shifted redward relative to the quasar’s systemic velocity. The continuum source associated with this emitter is detected in HST F140W imaging (Fig.~\ref{fig:regs}), supporting its classification as an independent galaxy within the halo environment with a redshift of $z=2.804 \pm 0.001$, based on the peak of the line emission. The consistency in spectral shape across regions suggests that Ly$\alpha$ emission in the halo follows a relatively uniform profile despite the kinematic and morphological complexity observed in the moment maps.

\subsection{VLBI radio observations and variability}\label{sec:vlbi-results}

\begin{figure}
\includegraphics[trim={1.5cm 0.3cm 2cm 0cm},clip,width=\columnwidth]{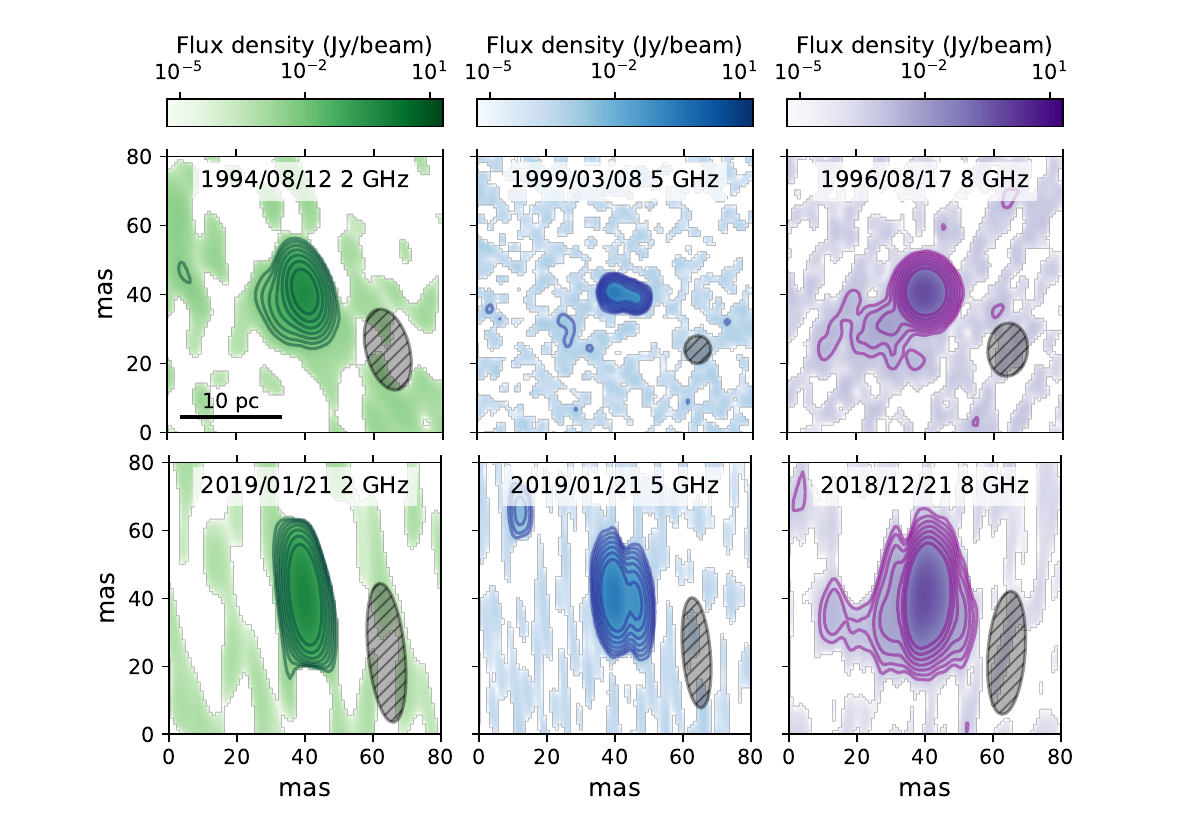}
\caption{Selection of VLBI images of J1017+6116 observed at various dates and frequencies: S band (2 GHz), X band (5 GHz), and C band (8 GHz), as labeled. The grayscale ellipses indicate the estimated PSF. The dark green, blue, and magenta contour lines correspond to flux density levels for the S, X, and C bands, calculated as multiples of the standard deviation ($3\times[1,\,2,\,4,\,8,\,16,\,32,\,64]$) above the mean background level. A complete set of VLBI observations is provided in Appendix~\ref{fig:VLBI}.}
\label{fig:VLBI_selected}
\end{figure}

\begin{figure}
\includegraphics[trim={0cm 0cm .3cm 0cm},clip,width=\columnwidth]{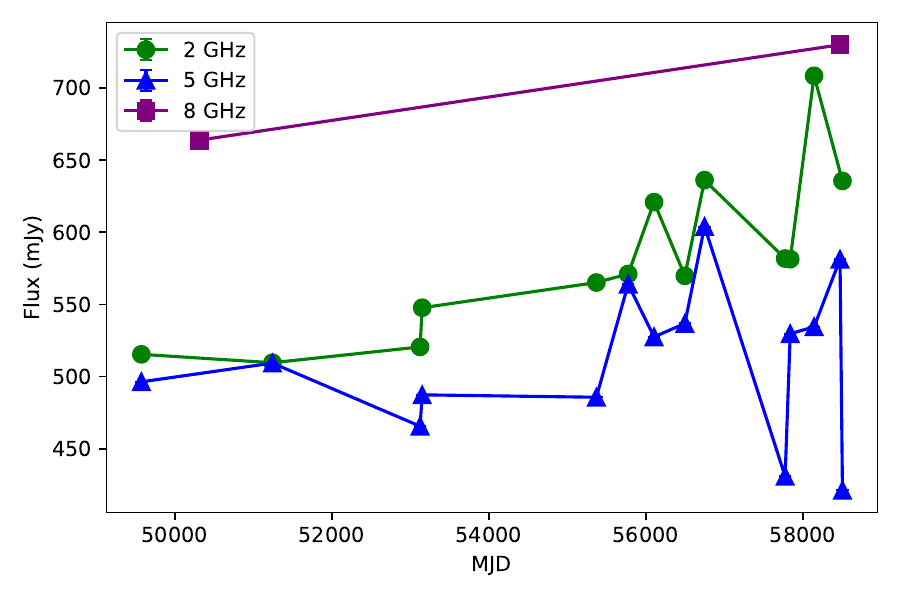}
\caption{Radio flux density measured over time at three different frequencies: 2 GHz (green circles), 5 GHz (blue triangles), and 8 GHz (purple squares). The data were obtained using VLBI observations and plotted against the modified Julian date (MJD). Error bars indicate uncertainties in the flux measurements at each frequency. The flux densities were calculated by integrating the emission above a 3-sigma threshold within the source region for each epoch and frequency. This shows the flux variability and evolution with time, with an overall increasing flux at 2 and 8~GHz.}
\label{fig:VLBI_time}
\end{figure}

Figure \ref{fig:VLBI_selected} presents a selection of VLBI images of J1017+6116 across three frequencies — S band (2 GHz), X band (5 GHz), and C band (8 GHz) — for multiple epochs. The contour lines in green, blue, and magenta correspond to different levels of flux density, calculated as multiples of the background standard deviation ($3\times\sigma$), with darker contours representing higher flux levels. The grayscale ellipses indicate the PSF for each observation. The images reveal compact radio structures near the quasar position, with variations in flux and morphology across different frequencies and epochs. No extended jet-like structure is detected at any of the observed frequencies. The complete set of VLBI images is shown in Fig.~\ref{fig:VLBI}.

The mean PA of $-136\pm3$ reported by \citet{2022ApJS..260....4P} aligns with the biconical feature observed in the Ly$\alpha$ halo. However, our reanalysis of the VLBI data did not find a consistent PA across different epochs or frequency bands. The only notable feature is a secondary radio component detected at 5 GHz (C band), located approximately 3.5 mas from the quasar, with a PA of $-108^\circ$ (see Fig. \ref{fig:VLBI_selected}, central panels). This secondary feature has a flux density about three times fainter than the primary source and remains present across all observed epochs. However, its nature remains unclear, and there is no clear evidence that it corresponds to a jet.

Figure \ref{fig:VLBI_time} presents the radio flux density of J1017+6116 over time at the three observed frequencies. The flux density measurements exhibit variability, with distinct fluctuations at all frequencies. The spatial extent of the radio emission remains confined to milliarcsecond (mas) scales throughout all epochs.

\section{Discussion}\label{sec:discussion}

\subsection{Quasar feedback and halo dynamics}\label{sec:qso-feedback}

Our results indicate that feedback from the quasar plays a crucial role in shaping the Ly$\alpha$ halo observed around CARLA J1017+6116. While RLAGNs are often associated with strong mechanical feedback from jets,  our observations suggest that both radiative and mechanical feedback mechanisms could be influencing the observed halo morphology. The radio variability of the source (Figs. \ref{fig:qso_sed} and \ref{fig:VLBI_time}), suggests episodic energy injection, which may impact ionization and gas dynamics within the CGM. However, it is also possible that the large-scale CGM itself is intrinsically asymmetric in density and structure, independent of direct AGN-driven feedback. Variations in gas density across different regions of the halo — resulting from past mergers, filamentary accretion, or previous episodes of AGN activity — could naturally lead to asymmetric Ly$\alpha$ emission and velocity dispersion. In this scenario, the observed Ly$\alpha$ morphology would be shaped not only by quasar feedback but also by preexisting CGM density variations that affect ionization, recombination, and scattering processes.

\subsection{Radiative feedback and the Ly$\alpha$ halo}

Radiative feedback from the quasar likely plays a significant role in shaping the Ly$\alpha$ emission in this system. The bright Ly$\alpha$ halo surrounding the quasar is consistent with strong photoionization of the CGM, where quasar radiation ionizes neutral hydrogen and enhances recombination line emission. This mechanism can drive large-scale ionization structures, regulate gas accretion, and shape the spatial distribution of emission \citep{2018MNRAS.477.1336C}. Unlike mechanical feedback, which depends on jet presence and orientation, radiative feedback is more isotropic and can extend over much larger scales.

However, the observed asymmetries in the Ly$\alpha$ halo suggest that, if radiative feedback is prevalent, the CGM gas distribution itself is likely inhomogeneous. Variations in neutral hydrogen column density, clumpy gas structures, or past interactions could create anisotropic ionization fronts despite the isotropic radiation field \citep[e.g.,][]{2016MNRAS.455.4100F,2012MNRAS.420..829O}. These asymmetries may influence the escape of ionizing photons, leading to preferential ionization in certain directions and shaping the observed halo morphology.

Additionally, the strong Ly$\alpha$ emission regions (R2 to R5) show complex kinematic structures that could result from radiative pressure on the surrounding medium. Photon momentum transfer can drive outflows and turbulence, contributing to the observed broad emission lines and asymmetric gas motions. If radiative feedback dominates, it suggests that powerful quasars can influence their CGM even in the absence of prominent jet-driven outflows.

The similarity of the Ly$\alpha$ spectral profiles across the halo, despite the variations in kinematics and morphology, suggests that ionization and radiative transfer effects are influencing the CGM in a relatively uniform manner. While the FWHM map indicates broad Ly$\alpha$ emission in regions R2, R4, and R5, the lack of strong secondary peaks or absorption features in the spectra implies that Ly$\alpha$ emission is predominantly shaped by large-scale quasar radiation rather than localized turbulent structures.

\subsection{Jet-induced shocks and gas motions}

The broadening of the Ly$\alpha$ emission lines in regions R2, R4, and R5 initially suggested the presence of jet-induced shocks propagating through the CGM, as broad Ly$\alpha$ line widths have been observed in systems where AGN-driven shocks interact with the surrounding gas \citep[e.g.,][]{2018MNRAS.479.1180M,2019A&A...632A..26G,2021A&A...650A..84V,2024MNRAS.528.4976D}. However, our reanalysis of the VLBI data did not reveal a persistent or well-collimated jet structure, making it difficult to attribute the observed gas motions and turbulence solely to jet-driven shocks. Instead, the observed radio variability and the shape of the radio component of the spectral energy distribution  (Figs.~\ref{fig:VLBI_time} and~\ref{fig:qso_sed}) suggests episodic energy injection, as seen in sources like 3C 454.3, where transient jet activity leads to fluctuations in radiative and kinetic energy deposition into the CGM \citep{1974ApJ...193...43B,2005AJ....130.1418J}. This variability can drive intermittent shocks and turbulence, influencing the observed kinematics without requiring a continuously extended jet.

An alternative explanation is that the jet is present but oriented close to our LoS, making extended emission difficult to detect in VLBI images. Projection effects, as seen in blazars and highly beamed quasars, can enhance core emission while suppressing large-scale lobes \citep[e.g.,][]{1995PASP..107..803U}. If the jet in CARLA J1017+6116 is aligned this way, significant radiative feedback could still be occurring, with episodic kinetic energy injection contributing to CGM turbulence and Ly$\alpha$ broadening.

The suppression of the jet may also be linked to its interaction with an inhomogeneous CGM, where turbulent gas and dense clumps disrupt its collimation and inhibit propagation \citep{2006ApJ...648L.101E,2022MNRAS.516..861T}. Simulations suggest that jets encountering dense clouds undergo Kelvin-Helmholtz instabilities, leading to turbulence and energy dissipation rather than efficient large-scale propagation \citep{2005MNRAS.359..781S}. These interactions could cause the jet to fragment or fully disrupt, preventing strong mechanical feedback, even if episodic jet activity has occurred in the past.

Alternatively, broad Ly$\alpha$ line widths in the CGM have also been linked to gravitational interactions and large-scale accretion shocks in merging environments \citep{2012ApJ...750...67R,2020MNRAS.492.1970G,2017MNRAS.466.3810F,2014ApJ...796..140P}. These processes could provide an alternative explanation for the observed kinematics, suggesting that multiple mechanisms — including prior jet activity, accretion shocks, and CGM turbulence — contribute to the observed Ly$\alpha$ halo structure.

\subsection{Quasar spectral features and their connection to large-scale halo processes}

An additional factor influencing the feedback mechanisms in CARLA J1017+6116 is the presence of a large DLA reported by \citet{2017MNRAS.472.1850G}. DLAs trace high-column-density neutral hydrogen in the CGM, often linked to inflows, outflows, or extended gas reservoirs around galaxies. The presence of such a dense CGM component could impact how quasar-driven feedback propagates, potentially altering the efficiency of both mechanical and radiative processes.

Despite the lack of strong, well-collimated jet signatures in the VLBI data, the presence of high-ionization UV lines such as C IV and O IV], along with the broad Ly$\alpha$ emission profile (Fig. \ref{fig:QSO_lines_spec}), suggests that radiative feedback is actively shaping the CGM. The extended red wing in the Ly$\alpha$ profile of the halo could be indicative of scattering effects, bulk gas motions, or even outflowing material. If jet activity was more prominent in the past, as suggested by the variability of the radio emission, then previous interactions between the quasar and the CGM may still influence the observed gas kinematics.

In this scenario, radiative feedback may dominate the current phase of quasar influence, with ionization-driven heating and outflows shaping the large-scale Ly$\alpha$ halo. However, past mechanical feedback from jet activity — potentially suppressed or redirected by interactions within the CGM — could have also played a role in establishing the observed structure. Further multiwavelength studies, particularly in X-ray and submillimeter regimes, will be essential to disentangle the relative contributions of radiative and mechanical processes in this system.

\subsection{The rarity of biconical Ly$\alpha$ halos in RLAGNs}

The distinct biconical shape of the Ly$\alpha$ halo in CARLA J1017+6116 raises the question of why such morphologies have not been commonly observed in other RLAGN systems. While radio jets are known to interact with the surrounding medium, producing ionized outflows and extended emission-line regions, our system appears to be one of the clearest examples of a biconical Ly$\alpha$ structure yet detected. Given that jet-driven feedback is expected to shape the CGM, the lack of similarly well-defined Ly$\alpha$ biconical halos in previous studies suggests that specific conditions must be met for such structures to form and remain detectable.

Several factors could contribute to this rarity. Projection effects and jet orientation relative to our LoS can obscure or distort an intrinsic biconical morphology, causing some systems to appear more circular or asymmetric in Ly$\alpha$ emission. Additionally, many RLAGNs exhibit episodic jet activity, with alternating phases of strong outflows and quiescence. If the biconical shape of the Ly$\alpha$ halo is linked to past jet activity, a system may only exhibit this morphology temporarily, depending on the timing of observations. Furthermore, the CGM is often clumpy and inhomogeneous, which can disrupt the expected outflow structure, preventing the formation of smooth, well-defined cones.

Another key factor is the interplay between radiative and mechanical feedback. In some cases, AGN radiation alone can carve out ionization cones, creating biconical morphologies even in the absence of a prominent jet. If radiative feedback dominates over mechanical effects, a strong alignment with the radio jet is not necessarily expected. Observational biases also play a role, as detecting well-defined biconical structures requires high-resolution and deep IFU surveys. Some systems may host similar morphologies, but without sufficient sensitivity, their Ly$\alpha$ halos may appear less structured. 

Finally, timescale differences between jet-driven kinematics and Ly$\alpha$ emission processes could further explain this misalignment. While jets operate on megayear timescales, Ly$\alpha$ emission, shaped by recombination and scattering, may persist for longer, meaning that the observed structure might not reflect the current jet phase but rather a past episode of AGN activity.
Understanding the specific conditions that lead to such well-defined biconical Ly$\alpha$ halos will require further investigation, including detailed comparisons with other RLAGNs and simulations that model both mechanical and radiative feedback effects.

\subsection{Connections with the CARLA J1017+6116 galaxy population}

\citet{2023A&A...670A..58M} provided a detailed analysis of the galaxy population in CARLA J1017+6116, showing that it hosts a high fraction of ETGs and enhanced or typical star formation activity in spectroscopically confirmed members with respect to star-forming galaxies on the main sequence at the same redshift \citep{2018ApJ...859...38N}.
The high ETG fraction points to early morphology transformation processes.

In close proximity the quasar, we spectroscopically confirm only one galaxy as a LAE, which is associated with the secondary Ly$\alpha$ halo. The secondary halo might indicate localized interactions between the quasar's feedback and the CGM of this nearby galaxy. However, the general absence of star-forming galaxies in the immediate environment of the quasar suggests that the feedback processes, particularly radiative feedback, has suppressed star formation on larger scales. The presence of a DLA in the quasar spectrum further supports this idea, indicating that significant amounts of neutral gas remain, but are likely affected by quasar-driven radiation or mechanical feedback.

The asymmetries in the Ly$\alpha$ halo and the kinematic complexity we observe suggest that quasar feedback is not only impacting individual galaxies but is also likely affecting the broader CGM and intracluster medium. 

\subsection{Comparison with other high-redshift halos}
\begin{figure}
\includegraphics[trim={0cm 0cm 1.8cm 0cm},clip,width=\columnwidth]{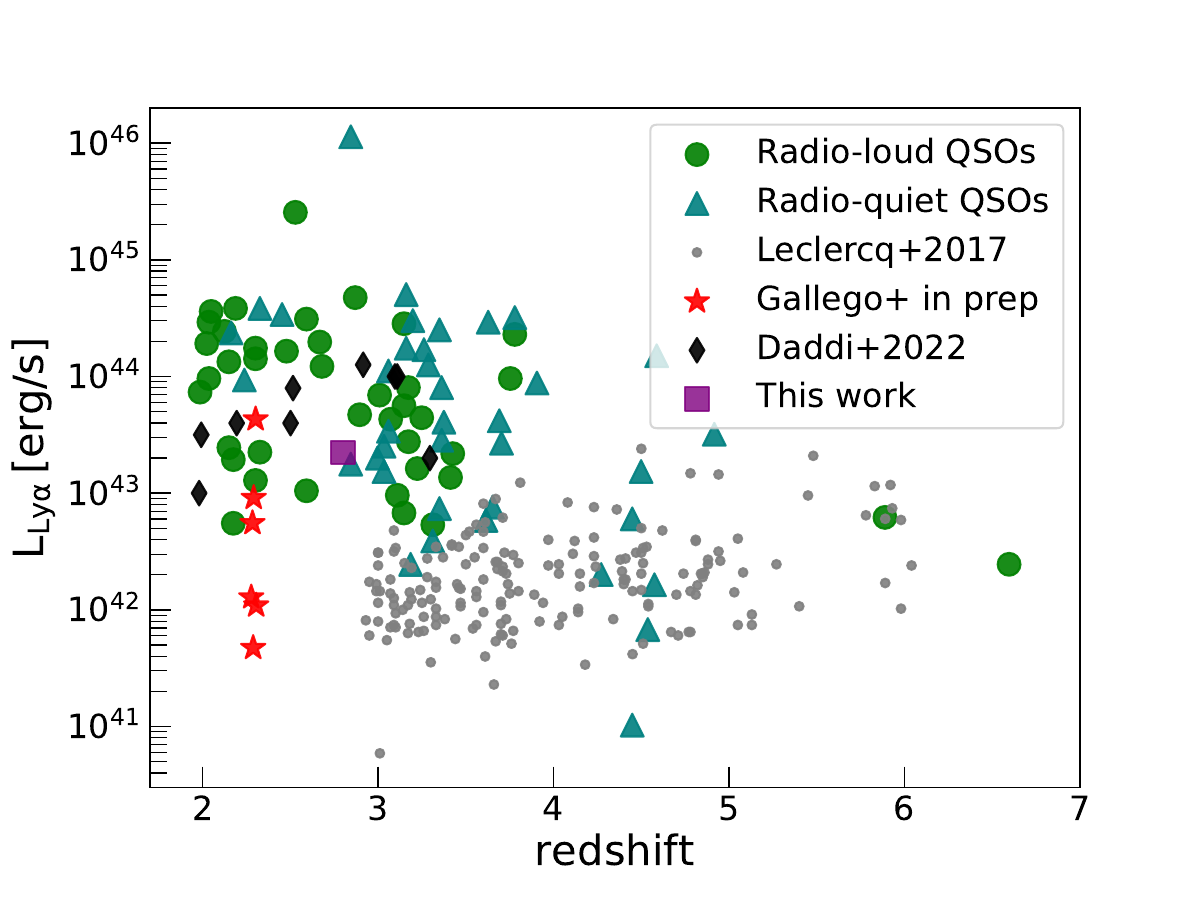}
\caption{Ly$\alpha$ halo luminosity as a function of redshift. The green circles and teal triangles correspond to Ly$\alpha$ halos detected around radio-loud and radio-quiet quasars, respectively. The gray dots correspond to Ly$\alpha$ halos around LAEs with MUSE data \citep{2017A&A...608A...8L}. The red stars correspond to the halos detected in the J0103+1316 protocluster at $z=2.3$ (Gallego \& Steidel, in preparation).}
\label{fig:lum-vs-z}
\end{figure}

The Ly$\alpha$ halo surrounding J1017+6116 remains one of the most extended and luminous halos observed at high redshift, particularly in association with a radio-loud quasar. Comparisons with similar systems in previous studies, such as \citet{2019MNRAS.482.3162A} and \citet{2022MNRAS.510..786S}, show that radio-loud quasars often host disturbed halos, suggesting feedback mechanisms at work. However, the lack of a persistent jet signature in our VLBI data implies that the large-scale asymmetries and dynamics observed in this system might not be fully explained by jet-driven feedback, as suggested in these previous studies.

As shown in Fig. \ref{fig:lum-vs-z}, we place the luminosity of our detected Ly$\alpha$ halo in the context of other known systems. The green circles and teal triangles represent Ly$\alpha$ halos detected around radio-loud and radio-quiet quasars, respectively, with the gray dots corresponding to Ly$\alpha$ halos around typical LAEs. Our detection, marked by the purple square, lies within the expected range for Ly$\alpha$ halos associated with quasars at $z \approx 2.8$, suggesting that the halo is relatively typical in terms of luminosity compared to other high-redshift systems. This implies that the feedback or environmental processes shaping the Ly$\alpha$ emission in this system are consistent with the broader population of quasar hosts, rather than indicating unusually strong or weak feedback mechanisms.

The presence of the DLA in the quasar spectrum is significant, as it points to substantial neutral gas still accreting onto the quasar. This dense neutral medium could be hindering the escape of a potential jet while still allowing radiative feedback to shape the CGM. This contrasts with other radio-loud systems, where more clear-cut evidence of jet-driven outflows is observed.

The recent study by \citet{2024ApJ...964...84S} examined the CGM of two radio-loud quasars, 3C 9 and 4C 05.84 at redshifts $z = 2.02$ and $z = 2.32$, respectively, using observations of Ly$\alpha$, He II, and C IV emission lines. Similar to J1017+6116, these quasars also show large-scale Ly$\alpha$ nebulae extending to about 100 kpc, along with spatially resolved, kinematically distinct nebulae of He II and C IV. However, in these cases, the CGM is found to extend into companion galaxies, highlighting the interaction between quasars and their immediate environment. For J1017+6116, while we observed a LAE with and extended halo in close proximity to the main halo, we did not detect comparably strong companion galaxy activity around the quasar. 

In addition, the study by \citet{2023A&A...680A..70W}, using MUSE data, revealed that high-redshift radio galaxies tend to have more asymmetric and disturbed Ly$\alpha$ nebulae than quasars, particularly radio-quiet ones. The alignment between radio jet PAs and the morphology of Ly$\alpha$ nebulae in high-redshift radio galaxies suggests that the orientation and presence of a jet significantly shape these structures. However, in CARLA J1017+6116, where no strong jet signature is detected, the asymmetry observed in the Ly$\alpha$ halo may be less dependent on jet orientation and more influenced by radiative processes and interactions with the CGM. Alternatively, if the jet activity is highly episodic, the current lack of strong mechanical feedback may simply reflect a quiescent phase in its duty cycle, rather than an absence of jet-driven influence altogether.

The enhanced or typical star formation activity observed in the spectroscopically confirmed members of the cluster surrounding CARLA J1017+6116 \citep{2018ApJ...859...38N} aligns with the strong ionization features reported by \citet{2024ApJ...964...84S}, suggesting that the quasar’s radiative feedback is contributing to enhanced star formation. However, our analysis identified only one galaxy in close proximity to the quasar, associated with the secondary Ly$\alpha$ halo. This apparent scarcity of nearby star-forming galaxies could be influenced by observational limitations, resolution constraints, or selection effects in our IFU observations and data analysis, rather than indicating a true absence of additional galaxies in the vicinity.

The study of J1017+6116 adds to the growing body of evidence that quasar feedback can manifest in various forms, including radiative and mechanical processes, and that these processes can be influenced by the specific conditions of the host environment. While mechanical feedback from jets may play a dominant role in shaping the CGM of some radio-loud quasars, the lack of strong jet signatures in CARLA J1017+6116 suggests that radiative feedback, given the potential suppression of mechanical outflows due to the dense interstellar medium, is the primary driver of the observed halo structure in this system.

\subsection{Future directions}

The discovery of a biconical Ly$\alpha$ halo in CARLA J1017+6116 highlights a strong connection between quasar feedback and CGM dynamics. However, its apparent rarity suggests that specific conditions — such as jet orientation, CGM structure, and the interplay between radiative and mechanical feedback — are necessary for such morphologies to emerge and persist. Future high-resolution studies, particularly with the James Webb space telescope (JWST) and the Atacama large millimeter/submillimeter array (ALMA), will be critical for assessing the prevalence of biconical Ly$\alpha$ halos and their connection to jet activity. A systematic survey of Ly$\alpha$-emitting halos in radio-loud quasars and high-redshift radio galaxies could determine whether these structures are transient phases of quasar evolution or a fundamental but rarely observed mode of AGN feedback.

To further disentangle the feedback mechanisms shaping CARLA J1017+6116, multiwavelength observations at higher spatial resolution will be essential. Infrared and submillimeter data from ALMA and JWST can trace dust-obscured star formation and molecular gas reservoirs, offering insights into the cold gas component and its role in fueling or regulating quasar activity. X-ray observations with \textit{Chandra} or \textit{XMM-Newton} could reveal the hot, ionized phase of the CGM, providing a clearer picture of the energy distribution and thermal state of the surrounding gas. Additionally, deeper and higher-resolution VLBI studies, complemented by multifrequency radio observations, could help clarify the nature of the secondary radio component and assess whether episodic jet activity played a role in shaping the CGM. 

By combining these observational efforts, we can develop a more comprehensive understanding of how quasar feedback — both radiative and mechanical — operates in high-redshift environments and influences large-scale galaxy evolution.

\section{Conclusions}\label{sec:conclusions}

In this paper we have presented a detailed analysis of the extended Ly$\alpha$ halo around the radio-loud quasar in the cluster CARLA J1017+6116 at $z=2.8$. Our study reveals significant morphological and kinematic complexities within the Ly$\alpha$ halo, with distinct asymmetries and broad spectral features indicative of interactions between the quasar and its surrounding environment. The halo extends at least 128 kpc in size and displays prominent SB enhancements and a broadening of Ly$\alpha$ lines in specific regions, with an overall biconical shape, which indicates that dynamic processes are at play.

While the initial analysis suggested a strong connection between the quasar’s radio emission and the Ly$\alpha$ halo biconical features, our reanalysis of the VLBI data introduces considerable uncertainties regarding the role of the quasar's radio variability. The PAs provided by previous studies were not confirmed with sufficient clarity in our reanalysis, making it difficult to conclude that the radio source or jet is directly responsible for shaping the large-scale Ly$\alpha$ halo. Instead, the small spatial scale of the radio variability (on the order of milliarcseconds) compared to the vast extent of the halo suggests that any mechanical feedback, if present, is localized or episodic. 

The discovery of a secondary radio component approximately 3.5 milliarcseconds from the quasar, about three times fainter than the primary source, raises additional questions. This secondary feature could represent episodic quasar activity or interactions with the interstellar medium, possibly linked to the DLA detected in the system. The presence of neutral gas indicates that quasar feedback, particularly radiative processes, could still be influencing the CGM and contributing to the observed Ly$\alpha$ halo features.

 \citet{2023A&A...670A..58M} show that galaxies in this cluster exhibit a high fraction of ETGs, while spectroscopically confirmed members experience enhanced or typical star formation activity compared to the main sequence at the same redshift. Although we did not find a physical connection between the quasar feedback processes and galaxy activity, a combination of radiative and mechanical effects may be influencing the broader cluster environment and may be contributing to the ongoing morphological transformation. The limited number of detected star-forming galaxies near the quasar may indicate localized suppression of star formation in the quasar’s immediate vicinity.

Our study highlights the complex interplay between quasar feedback and the CGM in a dense environment. The observed Ly$\alpha$ halo morphology and kinematics reflect the impact of both radiative and potential mechanical feedback from the quasar. However, the scale and variability of the radio emission introduce challenges in attributing the large-scale halo structure directly to a jet-driven mechanism. Future high-resolution, multiwavelength observations are needed to further explore the nature of this feedback and its role in galaxy evolution within high-redshift dense environments like CARLA J1017+6116.

\section*{Data availability}
The data underlying this article will be shared on reasonable request to the corresponding author.

\begin{acknowledgements}
This study was supported by the LabEx UnivEarthS, ANR-10-LABX-0023 and ANR-18-IDEX-0001. We are grateful to the referee for a constructive report. 
We acknowledge the use of the Very Long Baseline Array (VLBA) operated by the National Radio Astronomy Observatory (NRAO), which is a facility of the National Science Foundation operated under cooperative agreement by Associated Universities, Inc. The data used in this research were retrieved from the VLBA public archive. We also acknowledge the use of VLBA data under the US Naval Observatory's time allocation, supporting ongoing research into the celestial reference frame and geodesy.

Additionally, we acknowledge the use of data from the European VLBI Network (EVN), a joint facility of independent European, African, Asian, and North American radio astronomy institutes, and from the Long Baseline Array (LBA), part of the Australia Telescope National Facility, funded by the Commonwealth of Australia for operation as a National Facility managed by CSIRO.

We further acknowledge the use of data from the Korean VLBI Network (KVN) and the VLBI Exploration for Astrometry (KaVA), operated by the Korea Astronomy and Space Science Institute and the National Astronomical Observatory of Japan.

The VLBI images used in this study, including those from different epochs and frequencies (S-band, X-band, C-band), were analyzed by various experts, and we specifically acknowledge the following analysts: Yuri Y. Kovalev (yyk), Alexandr Pushkarev (pus), Greg Taylor (tay), and Leonid Petrov (pet). Their contributions to the VLBI FITS image database, including maps and visibility files, were instrumental to this work. We thank the Astrogeo VLBI FITS image database for providing access to these images.

This research made use of the Radio Fundamental Catalog (Petrov \& Kovalev, 2024, arXiv:2410.11794, DOI: 10.25966/dhrk-zh08).

The authors thank Gabriele Pezzulli and Rocco Lico for their insightful discussions on the VLBI data and jet production mechanisms.

\end{acknowledgements}

\bibliographystyle{aa}
\bibliography{refs}

\clearpage
\onecolumn

\begin{appendix}

\section{Additional figures}
\vspace{0.5cm}

\begin{figure}[h!]
\centering
\includegraphics[trim={3cm 0.5cm 2.5cm 0cm},clip,width=\columnwidth]{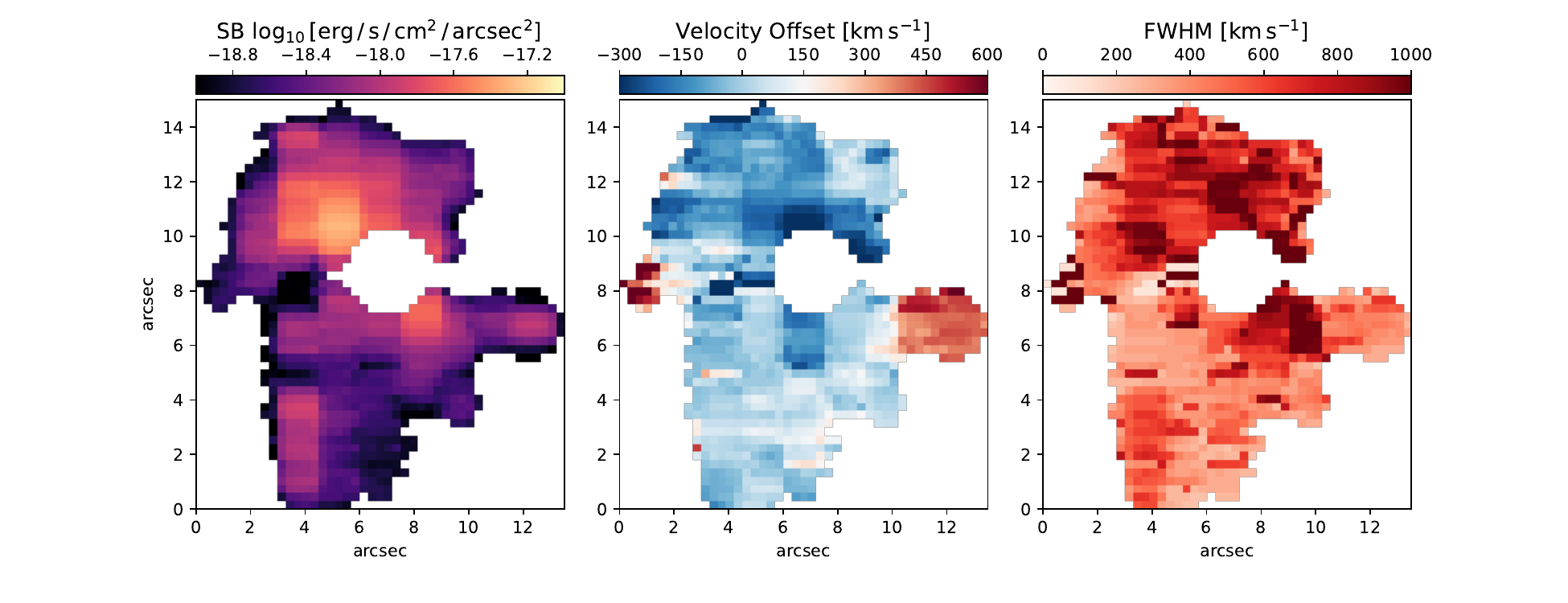}
\caption{Moment maps of the original data, using the same segmentation mask: SB (left), velocity offset (center), and FWHM (right).}
\label{fig:original_moments}
\end{figure}

\begin{figure}[h!]
\centering
\includegraphics[trim={0cm 0cm .3cm 0cm},clip,width=0.75\columnwidth]{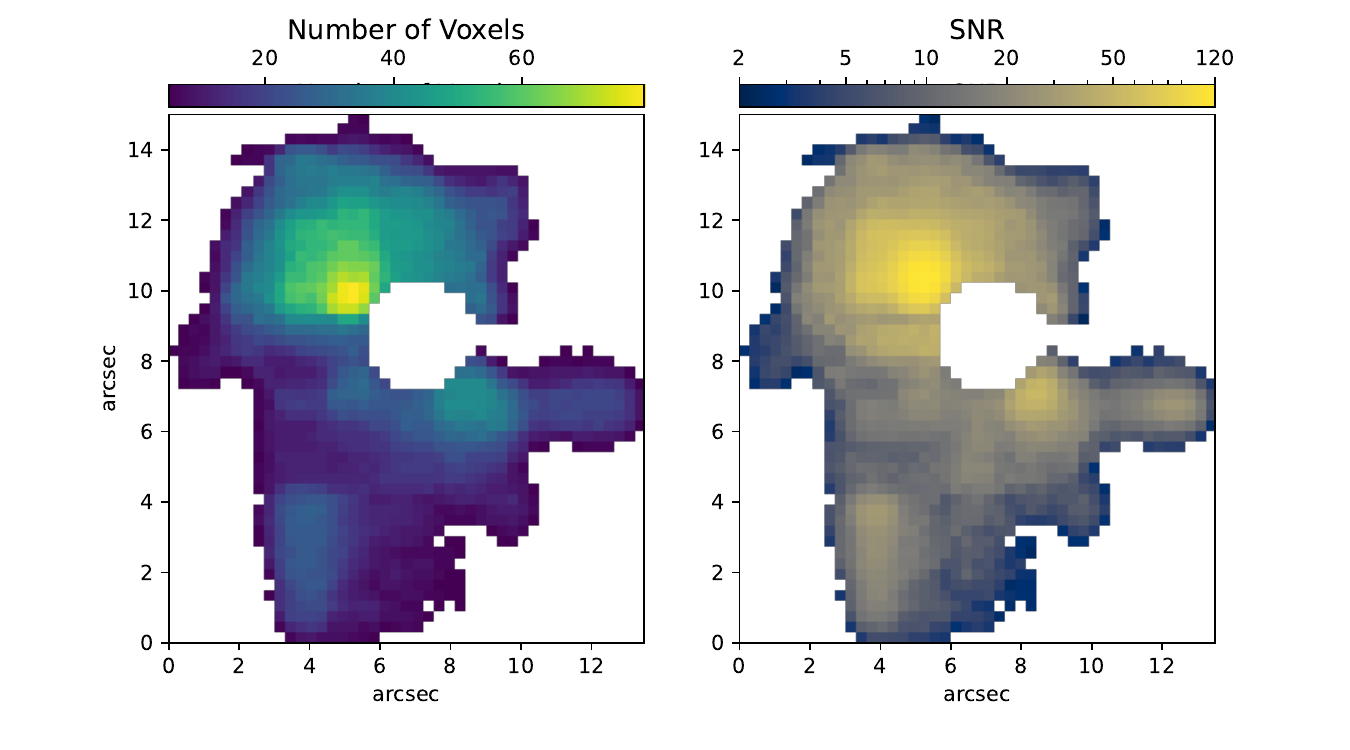}
\caption{Left: Number of voxels per pixel selected in the AKS segmentation mask. Only pixels with at least five connected voxels are included to ensure better spectral resolution. Right: S/N map of the AKS-detected Ly$\alpha$ emission, computed using the rescaled variance from the original unsmoothed datacube. The minimum displayed S/N is 2.3.}
\label{fig:SNR_Nvox}
\end{figure}

\begin{figure}[h!]
\centering
\includegraphics[trim={0cm 0cm .3cm 0cm},clip,width=\columnwidth]{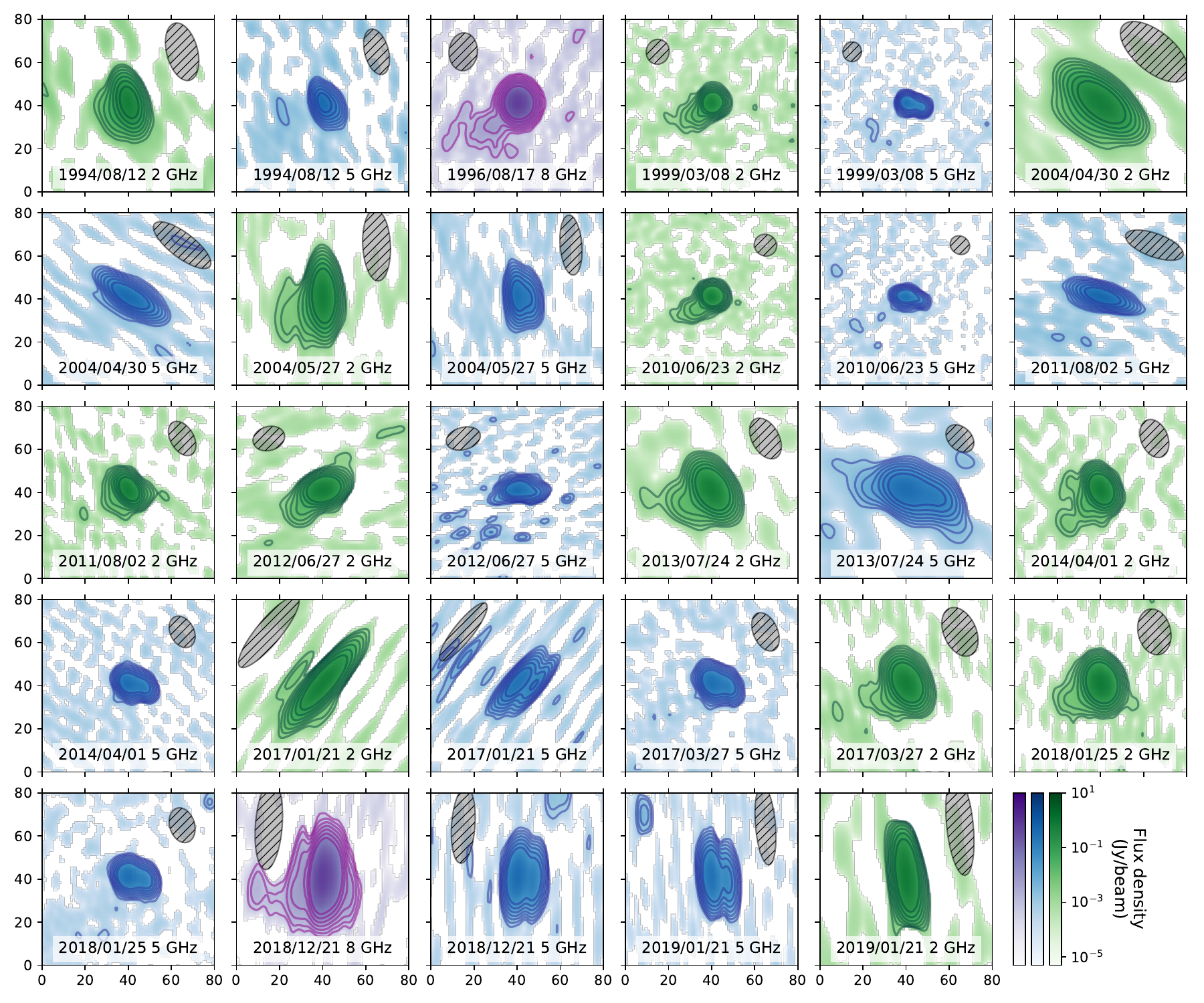}
\caption{VLBI images of J1017+6116 observed at various dates and frequencies: S band (2 GHz), X band (5 GHz), and C band (8 GHz), as labeled. The axis units are milliarcseconds to give the spatial extension of the regions. The grayscale ellipses represent the estimated PSF. The dark green, blue, and magenta contour lines indicate specific flux density levels for the S, X, and C bands, calculated as multiples of the standard deviation ($3\times[1,\,2,\,4,\,8,\,16,\,32,\,64]$) above the mean background level.}
\label{fig:VLBI}
\end{figure}

\end{appendix}

\end{document}